\documentclass[pra,twocolumn,superscriptaddress]{revtex4-2}
\usepackage{graphicx,empheq,amssymb,scalerel,tikz}

\usepackage[colorlinks=true, citecolor=blue, urlcolor=blue, linkcolor=blue ]{hyperref}
\usetikzlibrary{svg.path}
\definecolor{orcidlogocol}{HTML}{A6CE39}
\tikzset{
	orcidlogo/.pic={
		\fill[orcidlogocol] svg{M256,128c0,70.7-57.3,128-128,128C57.3,256,0,198.7,0,128C0,57.3,57.3,0,128,0C198.7,0,256,57.3,256,128z};
		\fill[white] svg{M86.3,186.2H70.9V79.1h15.4v48.4V186.2z}
		svg{M108.9,79.1h41.6c39.6,0,57,28.3,57,53.6c0,27.5-21.5,53.6-56.8,53.6h-41.8V79.1z M124.3,172.4h24.5c34.9,0,42.9-26.5,42.9-39.7c0-21.5-13.7-39.7-43.7-39.7h-23.7V172.4z}
		svg{M88.7,56.8c0,5.5-4.5,10.1-10.1,10.1c-5.6,0-10.1-4.6-10.1-10.1c0-5.6,4.5-10.1,10.1-10.1C84.2,46.7,88.7,51.3,88.7,56.8z};}}
\newcommand\orcid[1]{\href{https://orcid.org/#1}{\mbox{\scalerel*{\begin{tikzpicture}[yscale=-1,transform shape]\pic{orcidlogo};\end{tikzpicture}}{|}}}}
\begin{document}
\title{Long-Range Quantum Tunneling via Matter Waves}
\author{Yuan-Xing Yang}
\affiliation{School of Physical Science and Technology \& Lanzhou Center for Theoretical Physics, Lanzhou University, Lanzhou 730000, China}
\affiliation{Key Laboratory of Quantum Theory and Applications of MoE \& Key Laboratory of Theoretical Physics of Gansu Province, Lanzhou University, Lanzhou 730000, China}
\author{Si-Yuan Bai\orcid{0000-0002-4768-6260}}
\affiliation{School of Physical Science and Technology \& Lanzhou Center for Theoretical Physics, Lanzhou University, Lanzhou 730000, China}
\affiliation{Key Laboratory of Quantum Theory and Applications of MoE \& Key Laboratory of Theoretical Physics of Gansu Province, Lanzhou University, Lanzhou 730000, China}
\author{Jun-Hong An\orcid{0000-0002-3475-0729}}
\email{anjhong@lzu.edu.cn}
\affiliation{School of Physical Science and Technology \& Lanzhou Center for Theoretical Physics, Lanzhou University, Lanzhou 730000, China}
\affiliation{Key Laboratory of Quantum Theory and Applications of MoE \& Key Laboratory of Theoretical Physics of Gansu Province, Lanzhou University, Lanzhou 730000, China}

\begin{abstract}
Quantum tunneling is a quantum phenomenon in which a microscopic object crosses through a potential barrier even if its energy cannot overcome the barrier. A general belief is that tunneling occurs only when the barrier width is comparable to, or smaller than the de Broglie's wavelength of the object. Here, we study the tunneling of an ultracold atom among $N$ far-separated trapping potentials in a state-selective optical lattice and present a mechanism to realize long-range tunneling. We find that, mediated by the propagating matter wave emitted from the atom, coherent tunneling of the atom to the remote lattices occurs as long as bound states are present in the energy spectrum of the system formed by the atom and its matter wave. Going beyond the Markovian approximation, and breaking through the conventional distance constraint, our result opens another avenue to realizing tunneling and gives a guideline to developing tunneling devices.
\end{abstract}
\maketitle

\section*{Introduction} \label{introduction}
As one of the weirdest effects in microscopic world, quantum tunneling is a phenomenon consisting in a subatomic particle passing through a potential barrier with a potential energy greater than the energy of the particle \cite{RevModPhys.61.917,RevModPhys.66.217,RevModPhys.75.1,OLKHOVSKY2004133,WINFUL20061,Ankerhold2007QuantumTI,LANDSMAN20151,Zhu2024}.  It explains various radioactive decays of nuclei and how two nuclei overcome their mutual repulsion and fuse to generate huge energies \cite{RevModPhys.70.77,Hagino2012}. Quantum tunneling at metallic surfaces is the physical principle behind the operation of the scanning tunneling microscope \cite{PhysRevLett.49.57,PhysRevLett.50.1998,PhysRevLett.123.027402}. It also lays a foundation for nanotechnologies, such as the tunnel diode \cite{strambini2022superconducting} and resonant tunneling \cite{PhysRevLett.128.127701}, and some interdisciplinary sciences including quantum biology \cite{RevModPhys.35.724,doi:10.1126/sciadv.aaz4888,quantum3010006} and quantum chemistry \cite{doi:10.1126/science.1079294,wild2023tunnelling}. As a ubiquitous behavior of microscopic matters, such as electron \cite{suh2020electron}, proton \cite{PhysRevLett.112.148302}, nucleon \cite{RevModPhys.70.77}, photon \cite{PhysRevB.85.115425,PhysRevApplied.20.024057}, and superconducting Cooper pairs \cite{PhysRevLett.93.266803}, quantum tunneling has been observed in many experiments \cite{PhysRevLett.89.135505,PhysRevLett.98.263601,PhysRevLett.99.220403,PhysRevLett.100.190405,PhysRevLett.113.193003,wild2023tunnelling,PhysRevResearch.4.L012043}.

In parallel with its successful applications, the question of how long a particle takes to tunnel through a barrier has remained contentious since the early days of quantum mechanics \cite{RevModPhys.61.917,RevModPhys.66.217,RevModPhys.75.1,OLKHOVSKY2004133,WINFUL20061,Ankerhold2007QuantumTI,LANDSMAN20151,Zhu2024}. The progress on the ultracold-atom physics and attosecond science makes the direct measurement of the tunneling time possible \cite{PhysRevLett.117.010401,Foelling2007,doi:10.1126/science.1163439,Sainadh2019,PhysRevLett.119.023201,Ramos2020,Yu2022}. There is consensus on the fact that quantum tunneling is a consequence of the wave-particle duality and is more likely to occur when the width of the barrier is comparable to or smaller than the de Broglie wavelength of the particle. With the increase of the width of barriers, the tunneling rate experiences an exponential decrease \cite{PhysRevB.101.085410,yu2022highly}. Recently, how to realize a long-range tunneling has attracted much attention. For this purpose, the second-order quantum tunneling assisted by photons in the many-body lattice system \cite{PhysRevLett.107.095301,GallegoMarcos2015}, the virtual occupation of an intermediate site in the quantum-dot system \cite{Braakman2013}, and the molecular bridge in the chemical reaction paradigm \cite{Winkler2014,D3CS00662J} has been proposed. A higher-order resonant tunneling achieved by the interplay between the atomic interactions and the tilted force of the lattice can also realize a long-range tunneling up to five lattice sites \cite{doi:10.1126/science.1248402,Kessing2022}. The chaos in a periodically driven lattice system assists the long-range tunneling too \cite{PhysRevLett.126.174102}.  However, a common necessity of the above works is the nearest-neighbour hopping of the particle, which is equal to the overlap integral of the wave functions in the two sites. It indicates that they still do not essentially go beyond the distance constraint of the quantum tunneling.

\begin{figure}[tbp]
	\includegraphics[width=\linewidth]{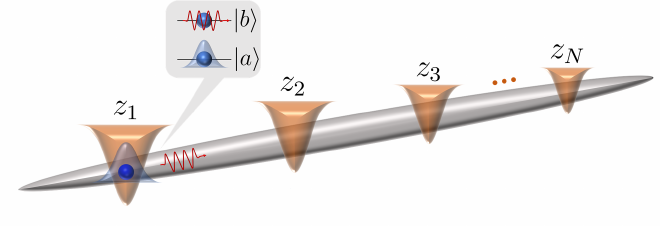}	
	\caption{\textbf{Scheme of the system.} An ultracold atom is confined in a state-selective optical lattice with different sites labelled by $z_j$ ($j=1,\cdots,N$) wrapped by an isolated tube. When it is in the ground state $|a\rangle$, the atom is trapped in the sites. When it is in the excited state $|b\rangle$, it propagates in the tube as a matter wave. The two states are coupled by a laser. The separation between the sites is so large that the atomic direct tunnelings among the sites are impossible. }\label{p1}
\end{figure}

Inspired by the experimental simulation of the atomic spontaneous emission by the matter wave of an ultracold atom in a state-selective optical lattice, which is dubbed ``quantum optics without photons'' \cite{krinner2018spontaneous,PhysRevResearch.2.043307,kwon2022formation}, we propose a scheme to realize a long-range tunneling of the atom among the lattice sites. The separation between the lattice sites is so large that the spatial wave functions of the internal ground-state atom confined in the sites have a physically negligible overlap and thus the direct tunneling cannot occur. By applying a laser, the atom jumps to the excited state and is converted into a propagating matter wave. Mediated by this matter wave, a long-range tunneling among the lattice sites occurs. It is interesting to find that such a tunneling sensitively depends on the features of the energy spectrum of the total system formed by the laser-controlled atom and its matter wave. As long as bound states out of the continuum (BOCs) or in the continuum (BICs) are present in the energy spectrum, a dynamically periodic tunneling, with frequencies equal to the differences of the energies of the bound states divided by $\hbar$, occurs among the lattice sites. This periodicity makes the tunneling time in our system well defined and experimentally measurable. Breaking through the conventional distance constraint on quantum tunneling, our result is beneficial to design quantum-tunneling devices.


\section*{Results and Discussion}
\subsection*{Model}\label{modl} The ultracold-atom system offers us with an ideal platform to simulate quantum optics and many-body physics \cite{PhysRevLett.101.260404,Navarrete_Benlloch_2011,PhysRevA.95.013626,PhysRevLett.122.203603,GonzalezTudela2018nonmarkovianquantum,bello2019unconventional,RevModPhys.80.885}.
Inspired by the experiments of the emission of atomic matter waves \cite{krinner2018spontaneous,PhysRevResearch.2.043307,kwon2022formation}, we investigate the tunneling dynamics of an ultracold atom in an optical lattice. Having two hyperfine states $|a\rangle$ and $|b\rangle$ and a mass $m$, the atom is confined in a state-selective optical lattice with $N$ unoccupied sites \cite{RevModPhys.80.885}. The lattice is embedded in an isolated tube, see Fig.~\ref{p1}. When the atomic internal state is $|a\rangle$, it is trapped in the spatial ground state $\Phi_a(z)=(\sqrt{\pi}\bar{z})^{-1/2}\exp[-z^2/(2\bar{z}^2)]$, with an eigenenergy $\tilde{\omega}/2$ and $\bar{z}=\sqrt{\hbar/(m\tilde{\omega})}$, by the lattice potential $V(z)=m\tilde{\omega}^2z^2/2$. Thus, the movement of the $|a\rangle$-state atom to other lattice sites is classically forbidden. Introducing an operator $\hat{\sigma}_j\equiv|0\rangle_j\langle 1|$, with $|0\rangle_j$ and $|1\rangle_j$ denoting the unoccupied and occupied states of the atom in the $j$th lattice site, we rewrite the Hamiltonian of the $|a\rangle$-state atom as $\hat{H}_{a}=\sum_{j=1}^N\hbar(\omega_a+\tilde{\omega}/2)\hat{\sigma}^\dag_j\hat{\sigma}_j$ with $\omega_a$ being the bare frequency of $|a\rangle$. On the contrary, when the atom is in the excited state $|b\rangle$ with a bare frequency $\omega_b$, it is set free from the site and propagates as a matter wave $\Phi_{b,k}(z)= e^{ikz}/\sqrt{L}$ along the tube, with $L$ being the tube length and $k=2\pi n/L$ ($n\in\mathbb{Z}$). Without feeling the optical lattice, the $|b\rangle$-state atom also has no chance to move to other lattice sites. The Hamiltonian of the $|b\rangle$-state atom is $\hat{H}_b=\sum_k\hbar[\omega_b+\hbar k^2/(2m)]\hat{b}^\dag_k\hat{b}_k$, where $\hat{b}_k$ is the annihilation operator of the $k$th mode of the emitted matter-wave quanta. The states $|a\rangle$ and $|b\rangle$ are coupled by a microwave field such that the atom performs the Rabi oscillation described by $\hat{H}_{ab}={\hbar\Omega\over 2}\sum_{j=1}^N\sum_{k}(\gamma_{jk}e^{i\omega_L t}\hat{\sigma}^\dag_j\hat{b}_k+\text{h.c.})$, where $\Omega$ and $\omega_L$ are the strength and the frequency of the driving field, and $\gamma_{jk}=\int dz\Phi_a(z-z_j)\Phi^*_{b,k}(z)$ is the Franck-Condon overlap between the trapped wave function at the $j$th site and the $k$th mode of the matter wave. The lattice sites are assumed to be so far separated that the overlap of the wave functions of the atom in different sites is much smaller than $|\gamma_{jk}|$ and thus the direct tunneling among them is negligible. This endows a substantial difference of our scheme from the ones in Refs. \cite{PhysRevLett.107.095301,GallegoMarcos2015,Braakman2013,Winkler2014,D3CS00662J,doi:10.1126/science.1248402,Kessing2022,PhysRevLett.126.174102}, where the nearest-neighbour hopping is necessary.  In a rotating frame with $\hat{H}_0=\sum_j\hbar(\omega_b-\omega_L)\hat{\sigma}_j^\dag\hat{\sigma}_j+\sum_k\hbar\omega_b\hat{b}_k^\dag\hat{b}_k$, the total Hamiltonian becomes \cite{PhysRevA.105.023703}
\begin{equation}
{\hat{H}\over\hbar}=\omega_0\sum_{j=1}^N\hat{\sigma}^\dag_j\hat{\sigma}_j+\sum_k[\omega_k\hat{b}_k^\dag\hat{b}_k+\sum_{j=1}^N(g_{jk}\hat{\sigma}^\dag_j\hat{b}_k+\text{h.c.})],\label{hmdtn}
\end{equation}where $\omega_0=\omega_a+{\tilde{\omega}\over2}-\omega_b+\omega_L$, $\omega_k={\hbar k^2\over2m}$, and $g_{jk}={\Omega\gamma_{jk}\over2}$. Having a rotating-wave approximate form, Eq. \eqref{hmdtn} guarantees the number of the atom always to be one. It rules out the nonphysical scenarios that a propagating $|b\rangle$-state atom and a trapped $|a\rangle$-state atom are created and annihilated simultaneously, which, respectively, are described by $\hat{\sigma}^\dag_j\hat{b}_k^\dag$ and $\hat{\sigma}_j\hat{b}_k$. The single-site case of Eq. \eqref{hmdtn} has been experimentally realized in Refs. \cite{krinner2018spontaneous,PhysRevResearch.2.043307}. It is noted that Eq. \eqref{hmdtn} takes a same form as the Tavis-Cummings model with a multimode boson field \cite{PhysRev.170.379}. Here, the $N$ sites of the optical lattice plays the role of $N$ identical two-level systems, each being separately coupled with the field.

\subsection*{Tunneling dynamics}\label{dynm}
Considering the atom is initially in the first lattice site, we have $|\Psi(0)\rangle=|a_1,\{0_k\}\rangle$. Its evolved state can be expanded as $|\Psi(t)\rangle=\sum_{j=1}^Nc_j(t)|a_j,\{0_k\}\rangle+\sum_kd_k(t)|b,1_k\rangle$, where $|a_j,\{0_k\}\rangle$ denotes that the atom in the state $|a\rangle$ is confined in the $j$th lattice site and no wave-matter quanta is triggered in the isolated tube and $|b,1_k\rangle$ denotes that the atom in the state $|b\rangle$ propagates in the isolated tube as a matter-wave quanta in the $k$th mode. From the Schr\"{o}dinger equation, we derive the equation of motion satisfied by the column vector $\mathbf{c}(t)=\left(
  \begin{array}{ccccc}
    c_1(t) & \cdots & c_N(t) \\
  \end{array}
\right)^\text{T}$ formed by the probability amplitudes of the atom in the $N$ sites as
\begin{equation}
\dot{\textbf{c}}(t)+i\omega_0\textbf{c}(t)+\int_0^td\tau\mathbf{f}(t-\tau)\mathbf{c}(\tau)=0.\label{dnct}
\end{equation}
Here $\mathbf{f}(t-\tau)=\int_0^\infty d\omega e^{-i\omega(t-\tau)}\mathbf{J}(\omega)$ is a $N$-by-$N$ correlation-function matrix of the matter wave and the element of the matrix $\mathbf{J}(\omega)$ is defined as $J_{jl}(\omega)=\sum_k g_{jk}^*g_{lk}\delta(\omega-\omega_k)$ characterizing the correlated spectral density of the atom in the $j$th and $l$th lattice sites. Such a correlation indicates that, although a direct tunneling of the atom among different lattice sites is absent, the mediated tunneling via the $|b\rangle$-state matter wave is triggered by the Rabi oscillation. One can readily derive
\begin{equation}
J_{jl}(\omega)=J_{|j-l|}(\omega)={\Omega^2e^{-2\omega\over\tilde{\omega}}\over\sqrt{8\pi\tilde{\omega}\omega}}\cos\Big[\Big({2\omega\over \tilde{\omega}}\Big)^{1\over2}{z_j-z_l\over\bar{z}}\Big].
\end{equation}
Reflecting the memory effect, the convolution in Eq. \eqref{dnct} renders the tunneling dynamics non-Markovian. Its dominant role in the dynamics has been observed \cite{krinner2018spontaneous,kwon2022formation}.

In the special case of the small-$\Omega$ limit, we can make the Markovian approximation to Eq. \eqref{dnct} by replacing ${\bf c}(\tau)$ by ${\bf c}(t)$ and extending the upper limit of the time integral to infinity \cite{Davies1974,Duemcke1979}. This approximation is also applicable when the spectral density reduces to a constant. The obtained solution is
\begin{equation}
    {\bf c}_\text{MA}(t)=\exp[-({\pmb \kappa}+i\omega_0+i{\pmb \Delta})t]{\bf c}(0),\label{makf}
\end{equation}where ${\pmb \kappa}=\pi {\bf J}(\omega_0)$ and ${\pmb \Delta}=\mathcal{P}\int{{\bf J}(\omega)d\omega\over\omega_0-\omega}$, with $\mathcal{P}$ being the Cauchy's principal value. We see that $|{\bf c}_\text{MA}(t)|^2$ exponentially decays to zero due to the positivity of ${\pmb \kappa}$. Thus, the atom asymptotically relaxes to the tube as a matter wave and no tunneling to other lattice sites occurs in the long-time condition of the Markovian dynamics. An exception occurs when $\omega_0<0$, where ${\pmb \kappa}=0$ but ${\pmb \Delta}\neq 0$ and thus a lossless tunneling can happen via exchanging the virtual matter waves \cite{PhysRevA.87.033831}.

The strong driving invalidates the Markovian approximation. In the general non-Markovian case, Eq. \eqref{dnct} can only be solved numerically. However, its asymptotic form is solvable by the Laplace transform method. It converts Eq. \eqref{dnct} into $[s+i\omega_0+\tilde{\bf f}(s)]\tilde{\mathbf{c}}(s)=\mathbf{c}(0)$, where $\tilde{\bf f}(s)=\int_0^\infty d\omega{{\bf J}(\omega)\over s+i\omega}$. Using the Jordan decomposition ${\bf D}(s)={\bf V}_s^{-1}\tilde{\bf f}(s){\bf V}_s=\text{diag}[D_{1}(s),\cdots,D_N(s)]$, we obtain $\tilde{\mathbf{c}}(s)=\mathbf{V}_s\big[ s+i\omega _{0}+\mathbf{D}(s)\big] ^{-1}\mathbf{V}_s^{-1}\mathbf{c}(0)$. $\mathbf{c}(t)$ is calculated by making the inverse Laplace transform to $\tilde{\mathbf{c}}(s)$, which needs finding the poles of $\tilde{\mathbf{c}}(s)$ from ($\varpi =is$)
\begin{equation}\label{eigen-solution}
Y_{j}(\varpi)\equiv\omega _{0}-iD_j(-i\varpi)=\varpi,~(j=1,\cdots, N).
\end{equation}
It is found that the root $\varpi$ multiplied by $\hbar$ is just the eigenenergy of Eq. \eqref{hmdtn}. It indicates that the atomic tunneling dynamics described by ${\bf c}(t)$ is intrinsically determined by the features of the energy spectrum of the total system formed by the confined atom and its propagating matter wave. We find that Eq. \eqref{eigen-solution} has three types of roots. First, $Y_j(\varpi)$ is ill-defined in the region of $\varpi>0$ due to the poles in $D_j(-i\varpi)$ and thus Eq. \eqref{eigen-solution} has an infinite number of roots, which form a continuous energy band. Second, because $Y_{j}(\varpi)$ is a decreasing function in the region out of the continuum, i.e., $\varpi<0$, Eq. \eqref{eigen-solution} has an isolated root $\varpi^\text{boc}_j$ provided $Y_{j}(0)<0$. The eigenstate corresponding to $\hbar\varpi^\text{boc}_j$ is called a BOC. Third, a removable singularity $\varpi^\text{bic}_j$ existing for $D_j(-i\varpi)$ when $\omega_0=\varpi^\text{bic}_j+iD_j(-i\varpi_j^\text{bic})$ also fulfills Eq. \eqref{eigen-solution}. The eigenstate corresponding to $\hbar\varpi^\text{bic}_j$ is called a BIC. The formation of the BOC and BIC has profound consequences on the non-Markovian dynamics \cite{PhysRevResearch.1.023027,PhysRevA.104.042609,PhysRevApplied.17.034073,PhysRevA.106.062438,PhysRevB.106.115427}. According to the Cauchy's residue theorem and contour integration, we have \cite{PhysRevA.103.L010601}
\begin{equation}
\mathbf{c}(t)=\mathbf{Z}(t)+\int_{0}^{\infty}\frac{d\varpi }{2\pi }[\mathbf{\tilde{c}}(0^{+}-i\varpi)-\mathbf{\tilde{c}}(0^{-}-i\varpi )]e^{-i\varpi t},\label{sol}
\end{equation}
where $\mathbf{Z}(t)=\sum_{\alpha=\text{bic,boc}}\sum_{j=1}^N$Res$[\mathbf{\tilde{c}}(-i\varpi _{j}^{\alpha})]e^{-i\varpi _{j}^{\alpha}t}$, with $\text{Res}[\mathbf{\tilde{c}}(-i\varpi _{j}^{\alpha})]$ being the residue contributed by the $j$th bound state obtained from $Y_j(\varpi)=\varpi$, and the second term comes from the continuous energy band. Containing an infinite number of superposition components oscillating with time in continuously changing frequencies $\varpi$, the second term tends to zero in the long-time limit due to the out-of-phase interference. Thus, if the bound state is absent, then $\lim_{t\rightarrow\infty}\mathbf{c}(t)={\bf 0}$ characterizes a complete relaxation of the atom in the isolated tube, while if the bound states are formed, then $\lim_{t\rightarrow\infty}\mathbf{c}(t)=\mathbf{Z}(t)$ implies a stable tunneling of the atom among the lattice sites. The result reveals that we can achieve a persistent matter-wave mediated long-range tunneling of the atom among the optical lattice by manipulating the formation of the bound states even when the lattice sites are so separated that the direct tunneling cannot happen. It inspires that, parallel to the rapidly developing electromagnetic-wave-based waveguide QED \cite{RevModPhys.95.015002}, our matter-wave-based waveguide supplies another realization of the efficient interconnect among well separated spatial nodes \cite{PRXQuantum.2.017002}. Note that, besides the real-valued eigenenergies of the BIC and BOC, Eq. \eqref{eigen-solution} also has complex-value eigenvalues. They can be found by analytically extending the function to the second Riemann sheet \cite{Lonigro2022}. These complex-valued eigenenergies contribute unstable oscillations and thus do not survive in the long-time tunneling dynamics.

\begin{figure}[tbp]
\includegraphics[width=\columnwidth]{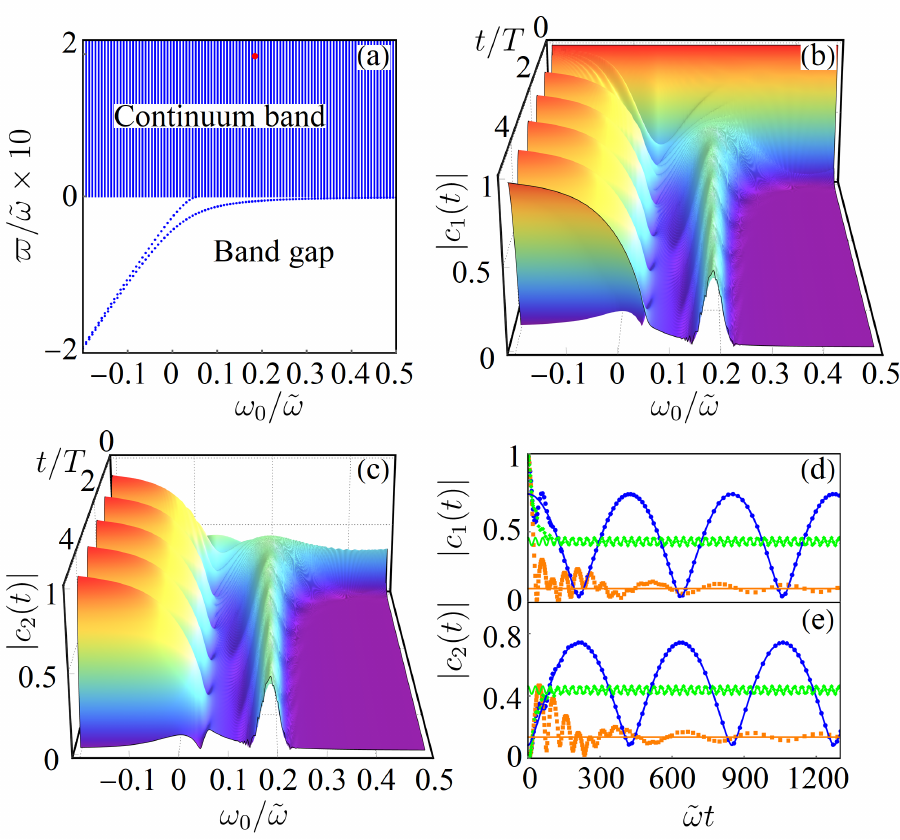}
\caption{\textbf{Energy spectrum and dynamics for different values of atomic frequency detuning in the case of two lattice sites.} \textbf{a} Energy spectrum and evolution of the absolute values of the probability amplitudes (\textbf{b}) $|c_1(t)|$ in the first site  and (\textbf{c}) $|c_2(t)|$ in the second site for different values of atomic frequency detuning $\omega_0$. The red dot in \textbf{a} marks the energy of the bound state in the continuum. $T$ in \textbf{b} and \textbf{c} defined as $2\pi/|\varpi_1^\text{boc}-\varpi_2^\text{boc/bic}|$ is the oscillation period in the steady state. Evolution of (\textbf{d}) $|c_1(t)|$ and (\textbf{e}) $|c_2(t)|$ when $\omega_{0}=-0.02\tilde{\omega}$ (blue dots), $0.06\tilde{\omega}$ (orange squares), and $0.18\tilde{\omega}$ (green rhombuses). Their long-time behaviors evaluated from Eq. \eqref{longC3} are shown by the solid lines in the same colors. We use $d=5\bar{z}$, $\Omega=0.13\tilde{\omega}$, $\bar{z}=0.065$ nm, and $\tilde{\omega}=2\pi\times 40$ kHz. }\label{dynamtwo}
\end{figure}

First, we assume that the lattice has two sites, i.e., $N=2$. It is easy to derive $D_{1/ 2}(s)=\tilde{f}_0(s)\pm\tilde{f}_1(s)$ and ${\bf V}_s=\begin{pmatrix} 1 & 1 \\ 1 & -1 \end{pmatrix}$. We thus have $\tilde{\bf c}(s)=\mathbf{M}[s+i\omega_0+\mathbf{D}(s)]^{-1}\begin{pmatrix} 1 & 1\end{pmatrix}^\text{T}$, where $\mathbf{M}=\begin{pmatrix} 1/2 & 1/2 \\ 1/2 & -1/2 \end{pmatrix}$. If the bound states are present, then
\begin{equation}\label{longC3}
c_j(\infty)= \sum_{\alpha\in{\text{bic,boc}}}\sum_{l=1}^2 M_{jl}Z_l^\alpha e^{-i\varpi_l^\alpha t},
\end{equation}
where $Z^\alpha_l=[1+\partial_sD_l(s)]^{-1}|_{s\rightarrow -i\varpi_l^\alpha}$. The $l$th BOC is formed when $Y_l(0)<0$. The divergence of $D_{1/2}(s)$ is removed by the BIC frequency determined by $J_0(\omega_{1/2}^\text{bic})\pm J_1(\omega_{1/2}^\text{bic})=0$ as
\begin{equation}
\varpi_{1/2}^\text{bic}=\tilde{\omega}(\bar{z}n_{1/2}\pi/d)^2/2,\label{bddcf}
\end{equation} with $d=|z_1-z_2|$, $n_1$ and $n_2$ being odd and even numbers, under the condition $\omega_0=\varpi_l^\text{bic}+iD_l(-i\varpi_l^\text{bic})$.
Equation \eqref{longC3} clearly shows that, in sharp contrast to complete relaxation to the isolated tube as a matter wave under the Markovian approximation in Eq. \eqref{makf}, a stable distribution and a persistently oscillating tunneling of the atom between the two sites occur when one and two bound states are formed, respectively. Such a tunneling mediated by the matter wave has not been reported before.

We plot in Fig. \ref{dynamtwo}a the energy spectrum of the system. It is found that two branches of BOCs are present and one BIC is formed at $\omega_0=0.18\tilde{\omega}$. The evolution of the modulus of the atomic probability amplitudes in the two lattice sites in Fig. \ref{dynamtwo}b,c shows that the atom completely relaxes into the isolated tube when no bound state is formed in the regime $\omega_0>0.25\tilde{\omega}$, which has no difference from the Markovian result in Eq. \eqref{makf}. In the regime $\omega_0\in (0.058,0.25)\tilde{\omega}$, one BOC is formed and thus both $|c_1(t)|$ and $|c_2(t)|$ tend to equal constants, see the orange squares in Fig. d, e. This corresponds to a stable distribution of the atom in the two lattice sites. An exception occurs when $\omega_0=0.18\tilde{\omega}$, where a BIC and a BOC coexists and thus the atom experiences a lossless oscillation between the two lattice sites in a frequency $|\varpi_1^\text{boc}-\varpi_2^\text{bic}|$, see the green rhombuses in Fig. \ref{dynamtwo}d, e. Such a lossless oscillation between the two lattice sites can also occur in a frequency $|\varpi_1^\text{boc}-\varpi_2^\text{boc}|$ when two BOCs are present in the regime $\omega_0<0.058\tilde{\omega}$, see the blue circles in Fig. \ref{dynamtwo}d, e. The matching of the numerical results with the analytic ones in Eq. \eqref{longC3} for the three typical parameter regimes confirms the dominant role of the bound states in the long-time dynamics. This lossless oscillation characterizes a coherent tunneling of the atom between the two sites mediated by the propagating matter wave. Thus, we achieve a long-range coherent tunneling even when the sites are so separated that the spatial wave functions of the atom have a negligible overlap.

\begin{figure}[tbp]
 \includegraphics[width=\columnwidth]{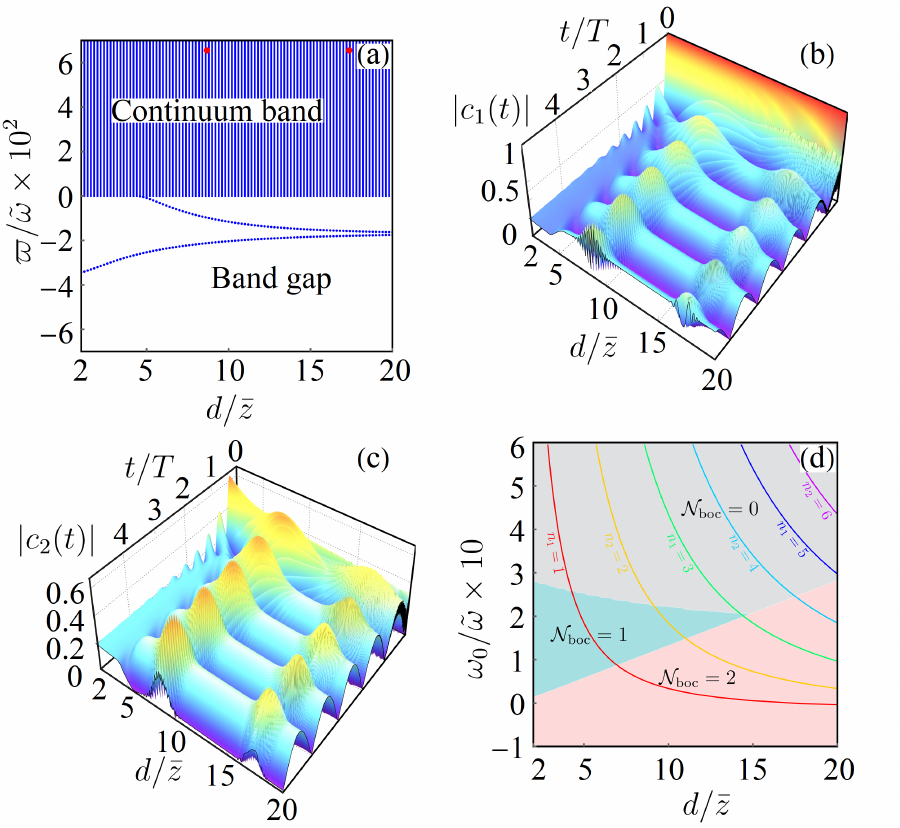}
 \caption{\textbf{Energy spectrum and dynamics for different values of the site separations in the case of two lattice sites.} \textbf{a} Energy spectrum and evolution of the absolute values of the probability amplitudes (\textbf{b}) $|c_1(t)|$ in the first site and (\textbf{c}) $|c_2(t)|$ in the second site for different values of the lattice-site separation $d$ when $\omega_0=0.05\tilde{\omega}$. Red dots in \textbf{a} mark the energies of the bound states in the continuum. \textbf{d} Phase diagram for forming different numbers of bound states out of the continuum denoted by $\mathcal{N}_\text{boc}$ in the $d$-$\omega_{0}$ plane. The solid lines evaluated from Eq. \eqref{bddcf} for different $n_l$ in \textbf{c} mark the positions forming the bound states in the continuum. Other parameters are the same as Fig. \ref{dynamtwo}. }\label{diffdd}
\end{figure}
Figure \ref{diffdd}a shows the energy spectrum of the system in different site separations $d$. Two branches of BOCs separate the spectrum into two regimes: one BOC when $d\le 5\bar{z}$ and two BOCs when $d>5\bar{z}$. A BIC is formed at $d=8.67\bar{z}$ and $17.36\bar{z}$, respectively. The evolution of $|c_j(t)|$ in Fig. \ref{diffdd}b, c indicates that $|c_1(t)|$ and $|c_2(t)|$ evolve to an equal constant in the single-BOC regime and experience a lossless oscillation with a common period $T=2\pi/|\varpi_1^\text{boc}-\varpi_2^\text{boc}|$ and a $\pi/2$ phase difference in the two-BOC regime. The latter reveals that the atom coherently goes back and forth between the two sites. With increasing $d$, $\varpi_1^\text{boc}$ tends closer and closer to $\varpi_2^\text{boc}$ and thus the period $T$ for the atom to finish one cyclic tunneling between the two sites becomes longer and longer. At the two distances in the presence of the BIC, $|c_j(t)|$ behaves a periodic oscillation with three frequencies, i.e., $|\varpi_1^\text{boc}-\varpi_2^\text{boc}|$ and $|\varpi_j^\text{boc}-\varpi^\text{bic}|$ ($j=1,2$). The phase diagram for forming different numbers of bound states in the $d$-$\omega_{0}$ plane is shown in Fig. \ref{diffdd}d. The formed zero, one, and two BOCs separate the diagram into three parts. Several discrete curves supporting the formation of the BIC in different $n_l$ are added above the diagram. It gives a global picture on the features of the atomic tunneling. If no bound state is present, then the atom relaxes as a matter wave in the tube. If one bound state is formed, then it tends to a stable equal-probability distribution in the two sites. If two or three bound states are formed, then the atom periodically and losslessly oscillates between the two sites in frequencies proportional to the eigenenergy difference of any pair of the bound states.

\begin{figure}
\includegraphics[width=\columnwidth]{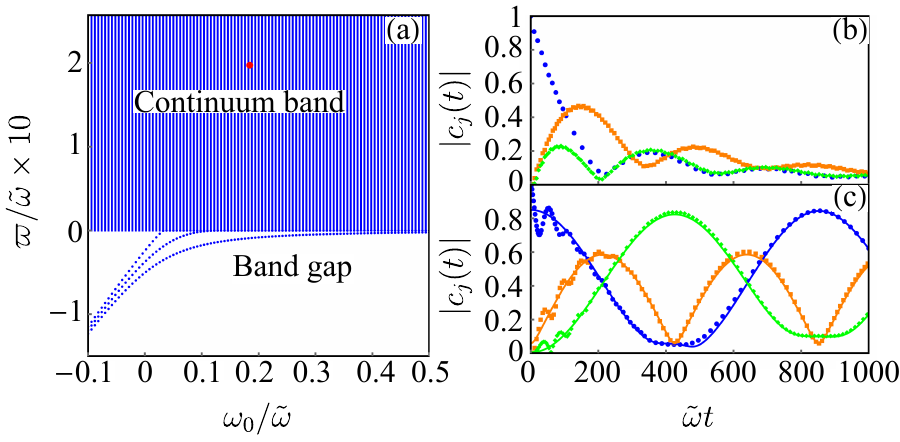}
\caption{\textbf{Energy spectrum and dynamics for different values of the atomic frequency detuning in the case of three lattice sites.} \textbf{a} Energy spectrum and evolution of the absolute values of the probability amplitudes $|c_1(t)|$ (blue dots), $|c_2(t)|$ (orange squares), and $|c_3(t)|$ (green rhombuses) when the atomic frequency detuning $\omega_0=0.4\tilde{\omega}$ in \textbf{b} and $-0.05\tilde{\omega}$ in \textbf{c} for the atom tunneling among three lattice sites. Red dot in \textbf{a} marks the energy of the bound state in the continuum. The long-time behaviors evaluated from Eq. \eqref{longCt3S} are shown by the solid lines in the same colors in \textbf{c}. We use $d=5\bar{z}$ and $\Omega=0.13\Tilde{\omega}$. }\label{Fig4}
\end{figure}

Our result can be generalized to the multiple-site case. When $N=3$, we have $D_1(s)=\tilde{f}_0-\tilde{f}_2$, $D_{2/3}(s)=\tilde{f}_0+{1\over2}(\tilde{f}_2\mp\tilde{e})$, and ${\bf V}_s=\begin{pmatrix} -1 & 1 & 1\\ 0 & {\tilde{f}_1(\tilde{e}-3\tilde{f}_2)\over\tilde{f}_2\tilde{e}-2\tilde{f}_1^2-\tilde{f}_2^2} &{\tilde{f}_1(\tilde{e}+3\tilde{f}_2)\over\tilde{f}_2\tilde{e}+2\tilde{f}_1^2+\tilde{f}_2^2}\\ 1&1&1\end{pmatrix}$, where the argument $s$ of $\tilde{f}_j(s)$ has been omitted for brevity and $\tilde{e}=(8\tilde{f}_1^2+\tilde{f}_2^2)^{1\over2}$. It is obtained that $\tilde{\bf c}(s)=\mathbf{M}[s+i\omega_0+\mathbf{D}(s)]^{-1}\begin{pmatrix} 1 & 1 & 1\end{pmatrix}^\text{T}$, where $\mathbf{M}=\begin{pmatrix} 1/2 & {1-\tilde{f}_2/\tilde{e}\over4} & {1+\tilde{f}_2/\tilde{e}\over4}\\ 0 & -{\tilde{f}_1/\tilde{e}} &{\tilde{f}_1/\tilde{e}}\\ -1/2&{1-\tilde{f}_2/\tilde{e}\over4}&{1+\tilde{f}_2/\tilde{e}\over4}\end{pmatrix}$. The BIC for $D_1(s)$ is formed at $\varpi_{1}^\text{bic}=\tilde{\omega}(\bar{z}n\pi/d)^2/8$, with $n$ being an even number. According to the residue theorem, we have
\begin{eqnarray}\label{longCt3Sdd}
c_j(\infty)=\sum_{\alpha=\text{bic,boc}}\sum_{l=1}^3{M_{jl}e^{st}\over 1+\partial_sD_l(s)}\Big|_{s\rightarrow-i\varpi_l^\alpha}.
\end{eqnarray}
The energy spectrum in Fig. \ref{Fig4}a shows that three BOCs and one BIC at most are formed. Without the bound state, the atom relaxes as a matter wave in the tube and no occupation in the sites survives, see Fig. \ref{Fig4}b. When three BOCs are present, $|c_1(t)|$ and $|c_2(t)|$ tend to a lossless oscillation in three frequencies $|\varpi_i^\text{boc}-\varpi_j^\text{boc}|$ ($i,j=1,2,3$), and $|c_2(t)|$ behaves as an oscillation in a frequency $|\varpi_2^\text{boc}-\varpi_3^\text{boc}|$, see Fig. \ref{Fig4}c. They match with our analytical result \eqref{longCt3Sdd} and verify the distinguished bound-state role in the tunneling dynamics. The result confirms again that we can realize a long-range tunneling among the lattice sites by the mediation of the matter wave.

\section*{Conclusions}\label{conclusion}
We have studied the tunneling dynamics of an ultracold atom in a state-selective optical lattice embedded in an isolated tube. A mechanism to realize a long-range tunneling of the atom among the lattice sites via the mediation of its propagating matter wave is found. In contrast to one's general belief that tunneling occurs when the atomic wavelength exceeds the width of the confined potential, our result reveals an alternative tunneling by converting the atom into a matter wave. We find that such a tunneling can be controlled by engineering the features of the energy spectrum of the system. When zero, one, and more BOCs or BICs are present in the energy spectrum, the tunneling exhibits complete relaxation in the tube, stable distribution, and lossless oscillation among the lattice sites, respectively. The observation of the bound-state effect in the single-lattice-site case \cite{krinner2018spontaneous,kwon2022formation} gives a support to realize our scheme. Supplying an insightful instruction to realize a controllable long-range tunneling, our result is helpful not only to simulate many-body physics with long-range orders \cite{RevModPhys.80.885,doi:10.1126/science.aal3837}, but also to design quantum-tunneling \cite{Yngvesson1991} and -interconnect \cite{PRXQuantum.2.017002,Niu2023} devices. In parallel to cavity \cite{Lei2023,Mirhosseini2019}, circuit \cite{Blais2020,RevModPhys.93.025005}, and waveguide QED systems \cite{PhysRevLett.132.090401,PhysRevLett.127.083602}, our matter-wave setup supplies a perfect realization of quantum interconnect among different quantum nodes. The matter-wave version  is advantageous over the photonic platforms because there is no loss for the matter waves.

\section*{Methods}
\subsection*{Derivation of dynamical evolution}
\begin{figure}
\includegraphics[angle=-90,width=0.45\columnwidth]{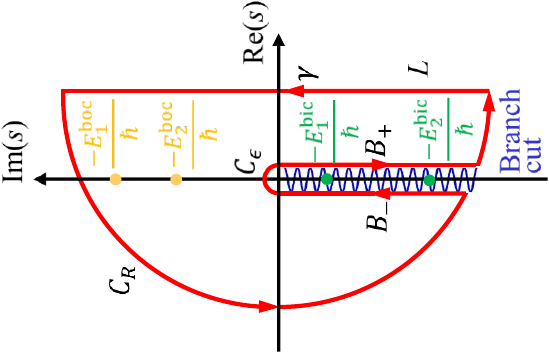}
\caption{\textbf{Contour integration to evaluate the inverse Laplace transform}. Path of the contour integration in the complex plane $\text{Re}(s)+i\text{Im}(s)$ for the calculation of the inverse Laplace transform of $\tilde{\textbf{c}}(s)$. The radii $C_R$ tends to infinity and $C_\epsilon$ tends to zero. The continuous poles form a branch cut wrapped by two inverse path $B_\pm$. The isolated poles form the bound states out of the continuum with eigenenergies $E^\text{boc}_j$ in the positive-Im($s$) axis and the bound states in the continuum with eigenenergies $E^\text{bic}_j$ in the negative-Im($s$) axis. }\label{smctit}
\end{figure}

The initial state of the total system is $|\Psi(0)\rangle=|a_1,\{0_k\}\rangle$. Its evolved state is expanded as $\Psi(t)\rangle=\sum_{j=1}^Nc_j(t)|a_j,\{0_k\}\rangle+\sum_kd_k(t)|b,1_k\rangle$. From the Schr\"{o}dinger equation, we derive
\begin{eqnarray}
i\dot{c}_{j}(t)&=&\omega_{0}c_{j}(t)+\sum_{k}g_{jk}d_{k}(t), \label{ct}\\
i\dot{d}_{k}(t)&=&\omega_{k}d_{k}(t)+\sum_{l=1}^Ng_{lk}^{*}c_{l}(t). \label{dkt}
\end{eqnarray}
Substituting the solution of Eq. \eqref{dkt} as $d_{k}(t)=-i\sum_{l}\int_{0}^{t}d\tau g_{lk}^{*}e^{-i\omega_{k}(t-\tau)}c_{l}(\tau)$ under the initial condition $d_k(0)=0$ into Eq. \eqref{ct}, we have
\begin{equation}
	\dot{c}_j(t)+i\omega_{0}c_{j}(t)+\sum_{l=1}^{N}\int_{0}^{t}d\tau f_{jl}(t-\tau)c_{l}(\tau)=0, \label{ct1}
\end{equation}
where $f_{jl}(t-\tau)=\int_0^\infty d\omega J_{jl}(\omega)e^{-i\omega(t-\tau)}$ is the correlation function of the matter wave and $J_{jl}(\omega)=\sum_kg_{jk}g_{lk}^*\delta(\omega-\omega_k)$ is the correlated spectral density of the atom between the $j$th and $l$th lattice sites. Remembering the form of $g_{jk}$ and $\gamma_{jk}$ and making the continuum limit of $k$ under $L\gg{\bar z}$, we obtain
\begin{eqnarray}
J_{jl}(\omega ) &=&\frac{\Omega ^{2}\sqrt{\pi }\bar{z}}{2L}\sum_{k}e^{ -\bar{z}%
^{2}k^{2}-ik(z_{l}-z_{j})} \delta (\omega -\omega _{k}) \nonumber\\
&=&{\Omega^2e^{-2\omega\over\tilde{\omega}}\over\sqrt{8\pi\tilde{\omega}\omega}}\cos\Big[\Big({2\omega\over \tilde{\omega}}\Big)^{1\over2}{z_j-z_l\over\bar{z}}\Big].\label{J_equationA}
\end{eqnarray}
We see that $J_{11}(\omega)=\dots=J_{NN}(\omega)$ and $J_{jl}(\omega)=J_{lj}(\omega)$. Thus, $J_{jl}$ satisfies the property $J_{jl}(\omega)=J_{|j-l|}(\omega)$. Equation \eqref{ct1} is rewritten as a column-vector form
\begin{equation}
\dot{\textbf{c}}(t)+i\omega_0\textbf{c}(t)+\int_0^td\tau\mathbf{f}(t-\tau)\mathbf{c}(\tau)=0,\label{smdnct}
\end{equation}
where $\mathbf{c}(t)=\left(
  \begin{array}{ccccc}
    c_1(t) & \cdots & c_N(t) \\
  \end{array}
\right)^\text{T}$, $\mathbf{f}(t-\tau)=\int_0^\infty d\omega e^{-i\omega(t-\tau)}\mathbf{J}(\omega)$ is a $N$-by-$N$ matrix with its elements being the correlation function $f_{jl}(t-\tau)$, and $\mathbf{J}(\omega)$ is a $N$-by-$N$ matrix with its elements being the correlated spectral density $J_{jl}(\omega)$.
Via a Laplace transform $\tilde{\mathbf{c}}(s)\equiv\int_{0}^{\infty }e^{-st}\mathbf{c}(t)dt$, Eq. \eqref{smdnct} is converted into
$[s+i\omega_0+\tilde{\bf f}(s)]\tilde{\mathbf{c}}(s)=\mathbf{c}(0)$, where $\tilde{\bf f}(s)=\int_0^\infty d\omega{{\bf J}(\omega)\over s+i\omega}$ is the Laplace transform of ${\bf f}(t-\tau)$.
Using the Jordan decomposition of $\tilde{\bf f}(s)$ as
\begin{equation}
{\bf D}(s)={\bf V}_s^{-1}\tilde{\mathbf{f}}(s){\bf V}_s=\text{diag}[D_{1}(s),\cdots,D_N(s)],\label{smjdd}
\end{equation}we have
\begin{eqnarray}
    \tilde{\bf c}(s)&=&{\bf V}_s[s+i\omega_0+{\bf D}(s)]^{-1}{\bf V}_s^{-1}\mathbf{c}(0)\nonumber\\
    &=&\mathbf{M}[s+i\omega_0+\mathbf{D}(s)]^{-1}\begin{pmatrix} 1 & \cdots & 1\end{pmatrix}^\text{T}.\label{smcssst}
\end{eqnarray}
The inverse Laplace transform reads $\mathbf{c}(t)=\int_{\gamma-i\infty}^{\gamma+i\infty}\frac{e^{st}}{2\pi i}\tilde{\bf c}(s)$, where $\gamma$ is chosen such that all the poles are included. According to the contour integration, see Fig. \ref{smctit}, the evaluation of the inverse Laplace transform needs finding the poles of $\tilde{\bf c}(s)$ from
\begin{equation}
s+i\omega_0+D_j(s)=0.\label{smpeqf}
\end{equation}
In general, Eq. \eqref{smpeqf} has three types of poles determined by $D_j(s)$, see Fig. \ref{smctit}. The first one is a continuum of poles along the negative imaginary axis of the complex plane formed by the real and imaginary parts of $s$, which forms a branch cut. The second one is several discrete roots formed by the removable singularities of $D_j(s)$ along the negative imaginary axis, which is called the BIC. The third one is the isolated poles along the positive imaginary axis, which is called the BOC.

According to the Cauchy's residue theorem and contour integration \cite{PhysRevA.103.L010601}, we have
\begin{equation}
\mathbf{c}(t)=\mathbf{Z}(t)+[\int_{B_+}\frac{d\varpi }{2\pi }+\int_{B_-}\frac{d\varpi }{2\pi }]\mathbf{\tilde{c}}(-i\varpi )e^{-i\varpi t},\label{sol}
\end{equation}
where $\mathbf{Z}(t)=\sum_{\alpha=\text{bic,boc}}\sum_{j=1}^N$Res$[\mathbf{\tilde{c}}(-i\varpi _{j}^{\alpha})]e^{-i\varpi _{j}^{\alpha}t}$, with $\text{Res}[\mathbf{\tilde{c}}(-i\varpi _{j}^{\alpha})]$ being the residue contributed by the $j$th bound state obtained from Eq. \eqref{smpeqf} with eigenenergy $E_j^\alpha=\hbar\varpi_j^\alpha$, and the second term comes from the branch cut of the contour. Oscillating with time in continuously changing frequencies, the second term tends to zero in the long-time limit due to the out-of-phase interference. Thus, if the bound state is absent, then $\lim_{t\rightarrow\infty}\mathbf{c}(t)={\bf 0}$ characterizes a complete relaxation of the atom in the isolated tube, while if the bound states are formed, then $\lim_{t\rightarrow\infty}\mathbf{c}(t)=\mathbf{Z}(t)$ implies either a stable distribution or a lossless oscillation of the atom among the optical lattice. Both of the results are absent in the Born-Markovian approximation.

\subsection*{Derivation of the energy spectrum}\label{smeigst}
The eigenstate of Eq. \eqref{hmdtn} is expanded as $|\Phi\rangle=\sum_{j=1}^N x_j|a_j,\{0_k\}\rangle+\sum_ky_k|b,1_k\rangle$. From the eigen equation $\hat{H}|\Phi\rangle=E|\Phi\rangle$, we obtain
\begin{eqnarray}
   (E/\hbar-\omega_0)x_j&=&\sum_kg_{jk}y_k,\label{smeix}\\
   (E/\hbar-\omega_k)y_k&=&\sum_{l=1}^Ng_{lk}^*x_l.\label{smeiy}
\end{eqnarray}
The substitution of the solution of Eq. \eqref{smeiy}, i.e., $y_k=\sum_{l=1}^Ng_{lk}^*x_l/(E/\hbar-\omega)$, into Eq. \eqref{smeix} leads to
$(E/\hbar-\omega_0)x_j=-i\sum_{l=1}^N\tilde{f}_{jl}(-iE/\hbar)x_l$, which can be tightly rewritten in a matrix form as
\begin{equation}
    [E/\hbar-\omega_0+i\tilde{\textbf{f}}(-iE/\hbar)]\textbf{x}=0,\label{smeige}
\end{equation}where $\textbf{x}=\left(
  \begin{array}{ccccc}
    x_1 & \cdots & x_N \\
  \end{array}
\right)^\text{T}$. The Jordan decomposition in Eq. \eqref{smjdd} readily convert Eq. \eqref{smeige} into
\begin{equation}
    E/\hbar-\omega_0+iD_j(-iE/\hbar)=0,\label{smegv}
\end{equation}which determines the eigenenergies $E$ of the system.

It is interesting to find that the pole equation \eqref{smpeqf} determining the tunneling dynamics is just the eigenenergy equation \eqref{smegv} after making the replacement of $E/\hbar=is$. Just for this reason, we call the eigenstates corresponding to the removable singularities of $D_j(s)$ the BICs and the ones corresponding to the the isolated poles the BOCs. This result reveals that the tunneling dynamics of the atom is essentially governed by the feature of the energy spectrum. If neither BOC nor BIC is formed, then $|c_j(t)|$ tends to zero with time and thus the atom eventually relaxes into the tube as a matter wave. If either the BOCs or BICs are formed, then the atom tends to a lossless coherent tunneling among the $N$ lattice sites in multiple frequencies determined by the difference of any pairs of the eigenenergies of the formed BOCs and BICs.

\section*{The case of $N=2$}
The spectral density matrix for the two lattice sites is ${\bf J}(\omega)=\left( \begin{array}{cc}
		J_0(\omega) & J_1(\omega) \\
		J_1(\omega) & J_0(\omega) \\
\end{array} \right)$. It contributes to the Laplace transform of $\textbf{f}(t-\tau)$ as $\tilde{\mathbf{f}}(s) =
	\left( \begin{array}{cc}
		\tilde{f}_0(s) & \tilde{f}_1(s) \\
		\tilde{f}_1(s) & \tilde{f}_0(s) \\
\end{array} \right)$, where $\tilde{f}_j(s)=\int_0^\infty d\omega{J_j(\omega)\over s+i\omega}$. The Jordan decomposition of $\tilde{\mathbf{f}}(s)$ reads \begin{equation}\label{2diagD}
{\bf D}(s)=\text{diag}[\tilde{f}_0(s)+\tilde{f}_1(s),\tilde{f}_0(s)-\tilde{f}_1(s)]
\end{equation}under ${\bf V}_s=\begin{pmatrix} 1 & 1 \\ 1 & -1 \end{pmatrix}$. Thus, we have
\begin{equation}
\tilde{\bf c}(s)=\left(
\begin{array}{cc}
1/2 & 1/2 \\
1/2 & -1/2%
\end{array}%
\right) \left(
\begin{array}{c}
\lbrack s+i\omega _{0}+D_{1}(s)]^{-1} \\
\lbrack s+i\omega _{0}+D_{2}(s)]^{-1}%
\end{array}%
\right).\label{smctt}
\end{equation}We need making the inverse Laplace transform to obtain ${\bf c}(t)$. The equations determining the poles of the inverse Laplace transform are
\begin{eqnarray}
    Y_1(\varpi)\equiv \omega_0-\int_0^\infty d\omega{J_0(\omega)+J_1(\omega)\over \omega-\varpi} =\varpi,\label{smy1}\\
    Y_2(\varpi)\equiv \omega_0-\int_0^\infty d\omega{J_0(\omega)-J_1(\omega)\over \omega-\varpi} =\varpi,\label{smy2}
\end{eqnarray}
where $\varpi=E/\hbar=is$. Their solutions can be classified in the following types.
\begin{enumerate}
    \item In the regime $\varpi>0$, the integration in both $Y_1(\varpi)$ and $Y_2(\varpi)$ is divergent. Thus, they jump rapidly between $\pm\infty$ and Eqs. \eqref{smy1} and \eqref{smy2} have infinite continuous roots, which forms an energy band after multiplying $\hbar$ as well as the branch cut.
    \item In the regime $\varpi<0$, both $Y_1(\varpi)$ and  $Y_2(\varpi)$ are a monotonically decreasing function with $\varpi$. Thus, one and only one root $\varpi_j^\text{boc}$ exists for either Eq. \eqref{smy1} or Eq. \eqref{smy2} as long as $Y_j(0)<0$. Since this type of root $\varpi_j^\text{boc}$ resides in a position out of the continuous energy band, we call its eigenstate BOC.
    \item In the regime $\varpi>0$, the integration in $Y_1(\varpi)$ and $Y_2(\varpi)$ has removable singularities $\varpi^\text{bic}_j$ determined by $J_0(\varpi^\text{bic}_{1})+J_1(\varpi^\text{bic}_1)=0$ and $J_0(\varpi^\text{bic}_2)-J_1(\varpi^\text{bic}_1)=0$ such that Eqs. \eqref{smy1} or \eqref{smy2} is respectively satisfied when $\omega_0=\varpi_1^\text{bic}+iD_1(-i\varpi_1^\text{bic})$ or $\omega_0=\varpi_2^\text{bic}+iD_2(-i\varpi_2^\text{bic})$. We readily evaluate that the removable singularities are $\varpi_{1/2}^\text{bic}=\tilde{\omega}(\bar{z}n_{1/2}\pi/d)^2/2$, with $d=|z_1-z_2|$, $n_1$ and $n_2$ being odd and even numbers. Since this type of root $\varpi_j^\text{bic}$ resides in a position in the continuous energy band, we call its corresponding eigenstate BIC.
\end{enumerate}
The residue contributed by the $l$th bound state with frequency $\varpi_l^\alpha$ evaluated from $Y_l(\varpi)=\varpi$ to the inverse Laplace transform of $[s+i\omega_0+D_j(s)]^{-1}$ in Eq. \eqref{smctt} is
\begin{eqnarray}
&&\lim_{s\rightarrow -i\varpi_l^\alpha}(s+i\varpi_l^\alpha)e^{st}[s+i\omega_0+D_j(s)]^{-1}\nonumber\\
&=&Z_l^\alpha e^{-i\varpi_l^\alpha t} \delta_{l,j},
\end{eqnarray}where $Z^\alpha_l=[1+\partial_sD_l(s)]^{-1}|_{s\rightarrow -i\varpi_l^\alpha}$.
Based on the fact that the steady-state solution of $\mathbf{c}(t)$ obtained via the inverse Laplace transform to $\tilde{\bf c}(s)$ is governed by the residues contributed by the eigenenergies of the BOCs and BICs, we readily have
\begin{eqnarray}
\mathbf{c}(\infty ) =\sum_{\alpha =\text{bic,boc}}
\left(
\begin{array}{cc}
1/2 & 1/2 \\
1/2 & -1/2%
\end{array}%
\right) \left(
\begin{array}{c}
Z_1^\alpha e^{-i\varpi
_{1}^{\alpha }t} \\
Z_2^\alpha e^{-i\varpi
_{2}^{\alpha }t}
\end{array}%
\right)
\end{eqnarray}in the case of two bound states are formed.

\begin{figure}[tbp]
	\includegraphics[width=\linewidth]{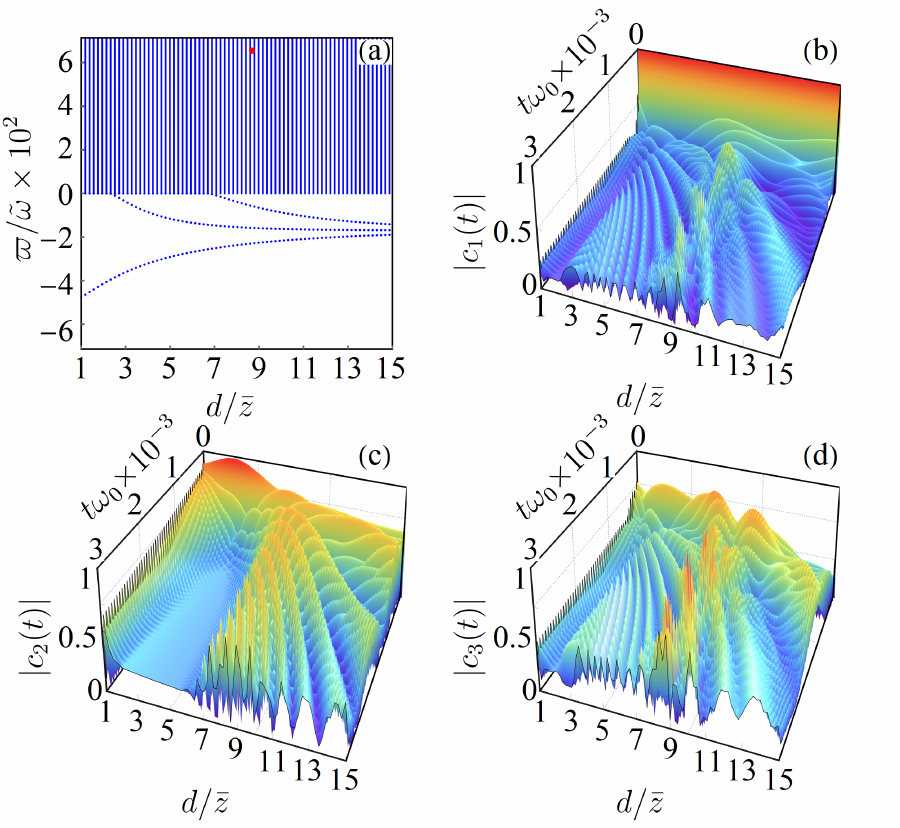}
	\caption{\textbf{Energy spectrum and dynamics for different values of the site separation in the case of three lattice sites.} \textbf{a} Energy spectrum for different values of the site separations $d$. The red point marks the position of the BIC. Evolution of the absolute values of the probability amplitudes (\textbf{b}) $|c_1(t)|$, (\textbf{c}) $|c_2(t)|$, and (\textbf{d}) $|c_3(t)|$. The parameters are $N=3$, $\Omega=0.13\tilde{\omega}$, and $\omega_{0}=0.05\tilde{\omega}$.}\label{FigEnergyS}
\end{figure}

\begin{figure}[tbp]
	\includegraphics[width=\linewidth]{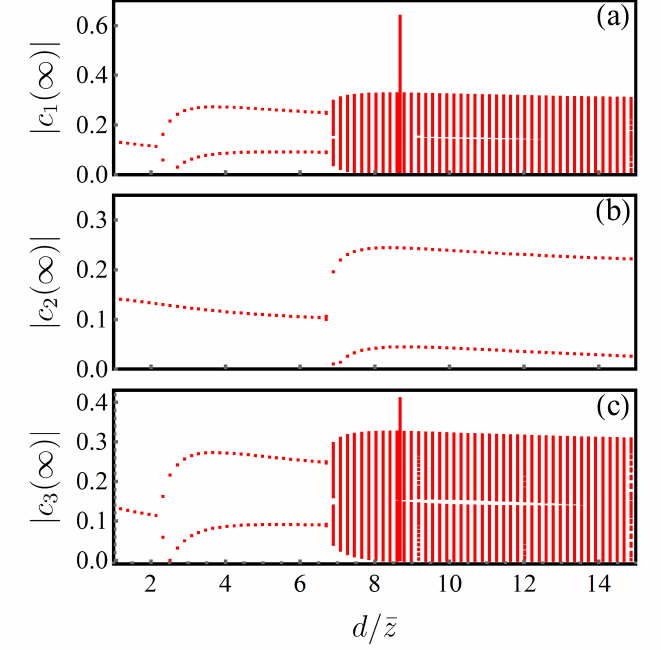}
	\caption{\textbf{Steady-state distribution of the atomic probability amplitudes in the three sites.} In the one-BOC regime, (\textbf{a}) $|c_{1}(\infty)|$, (\textbf{b}) $|c_2(\infty)|$, and (\textbf{c}) $|c_3(\infty)|$ are equal to constants. In the two-BOC regime, (\textbf{b}) $|c_2(\infty)|$ keeps to be a constant, while $|c_{1/3}(\infty)|$ in \textbf{a} and \textbf{c} exhibit periodic oscillations in single frequency, with the maxima and the minima marked by red dots. In the three-BOC regime, (\textbf{b}) $|c_{2}(\infty)|$ exhibit a single-frequency periodic oscillation, with the maxima and the minima marked by red dots, while $|c_{1/3}(\infty)|$ in \textbf{a} and \textbf{c} exhibit a lossless oscillation in three frequencies, with the local maxima and the minima joined by red lines. The parameters are the same as Fig. \ref{FigEnergyS}.  }\label{CLong}
\end{figure}

\section*{The case of $N=3$}
The spectral density matrix in the case of $N=3$ reads  ${\bf J}(\omega)=\left( \begin{array}{ccc}
		J_0(\omega) & J_1(\omega) & J_2(\omega)\\
		J_1(\omega) & J_0(\omega) & J_1(\omega) \\
        J_2(\omega) & J_1(\omega) & J_0(\omega) \\
\end{array} \right)$. It contributes to the Laplace transform of $\textbf{f}(t-\tau)$ as $\tilde{\mathbf{f}}(s) =\left(
\begin{array}{ccc}
\tilde{f}_{0} & \tilde{f}_{1} & \tilde{f}_{2} \\
\tilde{f}_{1} & \tilde{f}_{0} & \tilde{f}_{1} \\
\tilde{f}_{2} & \tilde{f}_{1} & \tilde{f}_{0}%
\end{array}%
\right)$, where the argument ``$s$'' of $\tilde{f}_j(s)$ has been omitted for brevity. The Jordan decomposition of $\tilde{\bf f}(s)$ is
\begin{eqnarray}
{\bf D}(s)=\text{diag}[\tilde{f}_0-\tilde{f}_2,\tilde{f}_0+{1\over2}(\tilde{f}_2-\tilde{e}),\tilde{f}_0+{1\over2}(\tilde{f}_2+\tilde{e})]~~~
\end{eqnarray}
under ${\bf V}_s=\begin{pmatrix} -1 & 1 & 1\\ 0 & {\tilde{f}_1(\tilde{e}-3\tilde{f}_2)\over\tilde{f}_2\tilde{e}-2\tilde{f}_1^2-\tilde{f}_2^2} &{\tilde{f}_1(\tilde{e}+3\tilde{f}_2)\over\tilde{f}_2\tilde{e}+2\tilde{f}_1^2+\tilde{f}_2^2}\\ 1&1&1\end{pmatrix}$, where $\tilde{e}=\sqrt{8\tilde{f}_1^2+\tilde{f}_2^2}$. We thus have $\tilde{\bf c}(s)$ in the case of $N=3$ as
$\tilde{\bf c}(s)=\mathbf{M}[s+i\omega_0+\mathbf{D}(s)]^{-1}\begin{pmatrix} 1 & 1 & 1\end{pmatrix}^\text{T}$, where $\mathbf{M}=\begin{pmatrix} 1/2 & {1-\tilde{f}_2/\tilde{e}\over4} & {1+\tilde{f}_2/\tilde{e}\over4}\\ 0 & -{\tilde{f}_1/\tilde{e}} &{\tilde{f}_1/\tilde{e}}\\ -1/2&{1-\tilde{f}_2/\tilde{e}\over4}&{1+\tilde{f}_2/\tilde{e}\over4}\end{pmatrix}$. The poles and the eigenenergies are determined by
\begin{equation}
    Y_j(\varpi)\equiv \omega_0-iD_j(-i\varpi)=\varpi. \label{smnet}
\end{equation}

Being similar to the $N=2$ case, Eq. \eqref{smnet} has three types of roots, i.e., continuous enegy band when $\varpi>0$, isolated poles $\varpi^\text{boc}_j<0$ provided $Y_j(0)<0$, and discrete removable singularities $\varpi^\text{bic}_j>0$. We can analytically determine that the roots of the BICs for $D_1(s)$ read $\varpi_1^\text{bic}=\tilde{\omega}(\bar{z}n\pi/d)^2/8$, with $n$ being an even number, which is formed when $\omega_0=\varpi_1^\text{bic}+iD_1(-i\varpi_1^\text{bic})$. Only the BOCs and BICs survives in the long-time dynamics. Thus, according to the residue theorem, we finally obtain
\begin{eqnarray}\label{longCt3S}
c_j(\infty)=\sum_{\alpha=\text{bic,boc}}\sum_{l=1}^3{M_{jl}e^{st}\over 1+\partial_sD_l(s)}\Big|_{s\rightarrow-i\varpi_l^\alpha}.
\end{eqnarray}
Equation \eqref{longCt3S} reveals that, in contrast to $c_1(\infty)$ and $c_3(\infty)$, $c_2(\infty)$ does not depend on $\tilde{f}_0(s)$. Therefore, the bound state formed by the $\tilde{f}_0(s)$-branch does not have an impact on $c_2(\infty)$.

The energy spectrum in Fig. \ref{FigEnergyS}a reveals that three BOCs and one BIC are formed at most. The evolution of $|c_j(t)|$ in Figs. \ref{FigEnergyS}b-d exhibits a one-to-one correspondence to the feature of the energy spectrum. When one BOC is present, the atom tends to a stable probability distribution in the three lattice sites. When two BOCs are present, $|c_{1/3}(t)|$ tend to a periodic oscillation in a frequency equal to the difference of the eigenenergies of the two BOCs divided by $\hbar$, while $|c_2(t)|$ still tends to be stable value. In this case, the emerged BOC is from $\tilde{f}_0$ and thus has no effect on $|c_2(t)|$. When three BOCs are present, $|c_1(t)|$ and $|c_3(t)|$ tend to an oscillation in three frequencies equal to the differences of the eigenenergies of any pairs of BOCs divided by $\hbar$, while $|c_2(t)|$ tends to a periodic oscillation in one frequency.
The analytical results as Eq. \eqref{longCt3S} are plotted in Fig. \ref{CLong}. The matching of our analytical results with the numerical ones verifies that the long-range tunneling is realized among the three lattice sites as long as multiple BOCs or BICs are present in the energy spectrum.

\begin{acknowledgments}
The work is supported by the National Natural Science Foundation of China (Grants No. 12275109, No. 12205128, and No. 12247101) and the Innovation Program for Quantum Science and Technology of China (Grant No. 2023ZD0300904).
\end{acknowledgments}
\textbf{Data availability}

The numerical data for generating the figures are available from authors upon request.

\textbf{Code availability}

The numerical codes for generating the results are also available from authors upon request.

\textbf{Author contributions}

Jun-Hong An proposed the original idea and developed the theoretical formulism. Yuan-Xing Yang and Si-Yuan Bai performed the analytic derivations. Yuan-Xing Yang drafted the manuscript and designed the figures. All authors discussed the results and contributed to the final manuscript.

\textbf{Competing interests}

The authors declare no competing interests.

\textbf{Correspondence} should be addressed to Jun-Hong An.

%


\begin{thebibliography}{81}%
\makeatletter
\providecommand \@ifxundefined [1]{%
 \@ifx{#1\undefined}
}%
\providecommand \@ifnum [1]{%
 \ifnum #1\expandafter \@firstoftwo
 \else \expandafter \@secondoftwo
 \fi
}%
\providecommand \@ifx [1]{%
 \ifx #1\expandafter \@firstoftwo
 \else \expandafter \@secondoftwo
 \fi
}%
\providecommand \natexlab [1]{#1}%
\providecommand \enquote  [1]{``#1''}%
\providecommand \bibnamefont  [1]{#1}%
\providecommand \bibfnamefont [1]{#1}%
\providecommand \citenamefont [1]{#1}%
\providecommand \href@noop [0]{\@secondoftwo}%
\providecommand \href [0]{\begingroup \@sanitize@url \@href}%
\providecommand \@href[1]{\@@startlink{#1}\@@href}%
\providecommand \@@href[1]{\endgroup#1\@@endlink}%
\providecommand \@sanitize@url [0]{\catcode `\\12\catcode `\$12\catcode
  `\&12\catcode `\#12\catcode `\^12\catcode `\_12\catcode `\%12\relax}%
\providecommand \@@startlink[1]{}%
\providecommand \@@endlink[0]{}%
\providecommand \url  [0]{\begingroup\@sanitize@url \@url }%
\providecommand \@url [1]{\endgroup\@href {#1}{\urlprefix }}%
\providecommand \urlprefix  [0]{URL }%
\providecommand \Eprint [0]{\href }%
\providecommand \doibase [0]{https://doi.org/}%
\providecommand \selectlanguage [0]{\@gobble}%
\providecommand \bibinfo  [0]{\@secondoftwo}%
\providecommand \bibfield  [0]{\@secondoftwo}%
\providecommand \translation [1]{[#1]}%
\providecommand \BibitemOpen [0]{}%
\providecommand \bibitemStop [0]{}%
\providecommand \bibitemNoStop [0]{.\EOS\space}%
\providecommand \EOS [0]{\spacefactor3000\relax}%
\providecommand \BibitemShut  [1]{\csname bibitem#1\endcsname}%
\let\auto@bib@innerbib\@empty
\bibitem [{\citenamefont {Hauge}\ and\ \citenamefont
  {St\o{}vneng}(1989)}]{RevModPhys.61.917}%
  \BibitemOpen
  \bibfield  {author} {\bibinfo {author} {\bibfnamefont {E.~H.}\ \bibnamefont
  {Hauge}}\ and\ \bibinfo {author} {\bibfnamefont {J.~A.}\ \bibnamefont
  {St\o{}vneng}},\ }\bibfield  {title} {\bibinfo {title} {Tunneling times: a
  critical review},\ }\href {https://doi.org/10.1103/RevModPhys.61.917}
  {\bibfield  {journal} {\bibinfo  {journal} {Rev. Mod. Phys.}\ }\textbf
  {\bibinfo {volume} {61}},\ \bibinfo {pages} {917} (\bibinfo {year}
  {1989})}\BibitemShut {NoStop}%
\bibitem [{\citenamefont {Landauer}\ and\ \citenamefont
  {Martin}(1994)}]{RevModPhys.66.217}%
  \BibitemOpen
  \bibfield  {author} {\bibinfo {author} {\bibfnamefont {R.}~\bibnamefont
  {Landauer}}\ and\ \bibinfo {author} {\bibfnamefont {T.}~\bibnamefont
  {Martin}},\ }\bibfield  {title} {\bibinfo {title} {Barrier interaction time
  in tunneling},\ }\href {https://doi.org/10.1103/RevModPhys.66.217} {\bibfield
   {journal} {\bibinfo  {journal} {Rev. Mod. Phys.}\ }\textbf {\bibinfo
  {volume} {66}},\ \bibinfo {pages} {217} (\bibinfo {year} {1994})}\BibitemShut
  {NoStop}%
\bibitem [{\citenamefont {van~der Wiel}\ \emph {et~al.}(2002)\citenamefont
  {van~der Wiel}, \citenamefont {De~Franceschi}, \citenamefont {Elzerman},
  \citenamefont {Fujisawa}, \citenamefont {Tarucha},\ and\ \citenamefont
  {Kouwenhoven}}]{RevModPhys.75.1}%
  \BibitemOpen
  \bibfield  {author} {\bibinfo {author} {\bibfnamefont {W.~G.}\ \bibnamefont
  {van~der Wiel}}, \bibinfo {author} {\bibfnamefont {S.}~\bibnamefont
  {De~Franceschi}}, \bibinfo {author} {\bibfnamefont {J.~M.}\ \bibnamefont
  {Elzerman}}, \bibinfo {author} {\bibfnamefont {T.}~\bibnamefont {Fujisawa}},
  \bibinfo {author} {\bibfnamefont {S.}~\bibnamefont {Tarucha}},\ and\ \bibinfo
  {author} {\bibfnamefont {L.~P.}\ \bibnamefont {Kouwenhoven}},\ }\bibfield
  {title} {\bibinfo {title} {Electron transport through double quantum dots},\
  }\href {https://doi.org/10.1103/RevModPhys.75.1} {\bibfield  {journal}
  {\bibinfo  {journal} {Rev. Mod. Phys.}\ }\textbf {\bibinfo {volume} {75}},\
  \bibinfo {pages} {1} (\bibinfo {year} {2002})}\BibitemShut {NoStop}%
\bibitem [{\citenamefont {Olkhovsky}\ \emph {et~al.}(2004)\citenamefont
  {Olkhovsky}, \citenamefont {Recami},\ and\ \citenamefont
  {Jakiel}}]{OLKHOVSKY2004133}%
  \BibitemOpen
  \bibfield  {author} {\bibinfo {author} {\bibfnamefont {V.~S.}\ \bibnamefont
  {Olkhovsky}}, \bibinfo {author} {\bibfnamefont {E.}~\bibnamefont {Recami}},\
  and\ \bibinfo {author} {\bibfnamefont {J.}~\bibnamefont {Jakiel}},\
  }\bibfield  {title} {\bibinfo {title} {Unified time analysis of photon and
  particle tunnelling},\ }\href
  {https://doi.org/https://doi.org/10.1016/j.physrep.2004.06.001} {\bibfield
  {journal} {\bibinfo  {journal} {Physics Reports}\ }\textbf {\bibinfo {volume}
  {398}},\ \bibinfo {pages} {133} (\bibinfo {year} {2004})}\BibitemShut
  {NoStop}%
\bibitem [{\citenamefont {Winful}(2006)}]{WINFUL20061}%
  \BibitemOpen
  \bibfield  {author} {\bibinfo {author} {\bibfnamefont {H.~G.}\ \bibnamefont
  {Winful}},\ }\bibfield  {title} {\bibinfo {title} {Tunneling time, the
  {H}artman effect, and superluminality: A proposed resolution of an old
  paradox},\ }\href
  {https://doi.org/https://doi.org/10.1016/j.physrep.2006.09.002} {\bibfield
  {journal} {\bibinfo  {journal} {Physics Reports}\ }\textbf {\bibinfo {volume}
  {436}},\ \bibinfo {pages} {1} (\bibinfo {year} {2006})}\BibitemShut {NoStop}%
\bibitem [{\citenamefont {Ankerhold}(2007)}]{Ankerhold2007QuantumTI}%
  \BibitemOpen
  \bibfield  {author} {\bibinfo {author} {\bibfnamefont {J.}~\bibnamefont
  {Ankerhold}},\ }\href {https://doi.org/10.1007/3-540-68076-4} {\emph
  {\bibinfo {title} {Quantum Tunneling in Complex Systems: The Semiclassical
  Approach}}}\ (\bibinfo  {publisher} {Springer Berlin, Heidelberg},\ \bibinfo
  {year} {2007})\BibitemShut {NoStop}%
\bibitem [{\citenamefont {Landsman}\ and\ \citenamefont
  {Keller}(2015)}]{LANDSMAN20151}%
  \BibitemOpen
  \bibfield  {author} {\bibinfo {author} {\bibfnamefont {A.~S.}\ \bibnamefont
  {Landsman}}\ and\ \bibinfo {author} {\bibfnamefont {U.}~\bibnamefont
  {Keller}},\ }\bibfield  {title} {\bibinfo {title} {Attosecond science and the
  tunnelling time problem},\ }\href
  {https://doi.org/https://doi.org/10.1016/j.physrep.2014.09.002} {\bibfield
  {journal} {\bibinfo  {journal} {Physics Reports}\ }\textbf {\bibinfo {volume}
  {547}},\ \bibinfo {pages} {1} (\bibinfo {year} {2015})}\BibitemShut {NoStop}%
\bibitem [{\citenamefont {Zhu}\ \emph {et~al.}(2024)\citenamefont {Zhu},
  \citenamefont {Tong}, \citenamefont {Liu}, \citenamefont {Yang},
  \citenamefont {Gong}, \citenamefont {Jiang}, \citenamefont {Lu},
  \citenamefont {Li}, \citenamefont {Song},\ and\ \citenamefont
  {Wu}}]{Zhu2024}%
  \BibitemOpen
  \bibfield  {author} {\bibinfo {author} {\bibfnamefont {M.}~\bibnamefont
  {Zhu}}, \bibinfo {author} {\bibfnamefont {J.}~\bibnamefont {Tong}}, \bibinfo
  {author} {\bibfnamefont {X.}~\bibnamefont {Liu}}, \bibinfo {author}
  {\bibfnamefont {W.}~\bibnamefont {Yang}}, \bibinfo {author} {\bibfnamefont
  {X.}~\bibnamefont {Gong}}, \bibinfo {author} {\bibfnamefont {W.}~\bibnamefont
  {Jiang}}, \bibinfo {author} {\bibfnamefont {P.}~\bibnamefont {Lu}}, \bibinfo
  {author} {\bibfnamefont {H.}~\bibnamefont {Li}}, \bibinfo {author}
  {\bibfnamefont {X.}~\bibnamefont {Song}},\ and\ \bibinfo {author}
  {\bibfnamefont {J.}~\bibnamefont {Wu}},\ }\bibfield  {title} {\bibinfo
  {title} {Tunnelling of electrons via the neighboring atom},\ }\href
  {https://doi.org/10.1038/s41377-023-01373-2} {\bibfield  {journal} {\bibinfo
  {journal} {Light Sci. Appl.}\ }\textbf {\bibinfo {volume} {13}},\ \bibinfo
  {pages} {18} (\bibinfo {year} {2024})}\BibitemShut {NoStop}%
\bibitem [{\citenamefont {Balantekin}\ and\ \citenamefont
  {Takigawa}(1998)}]{RevModPhys.70.77}%
  \BibitemOpen
  \bibfield  {author} {\bibinfo {author} {\bibfnamefont {A.~B.}\ \bibnamefont
  {Balantekin}}\ and\ \bibinfo {author} {\bibfnamefont {N.}~\bibnamefont
  {Takigawa}},\ }\bibfield  {title} {\bibinfo {title} {Quantum tunneling in
  nuclear fusion},\ }\href {https://doi.org/10.1103/RevModPhys.70.77}
  {\bibfield  {journal} {\bibinfo  {journal} {Rev. Mod. Phys.}\ }\textbf
  {\bibinfo {volume} {70}},\ \bibinfo {pages} {77} (\bibinfo {year}
  {1998})}\BibitemShut {NoStop}%
\bibitem [{\citenamefont {Hagino}\ and\ \citenamefont
  {Takigawa}(2012)}]{Hagino2012}%
  \BibitemOpen
  \bibfield  {author} {\bibinfo {author} {\bibfnamefont {K.}~\bibnamefont
  {Hagino}}\ and\ \bibinfo {author} {\bibfnamefont {N.}~\bibnamefont
  {Takigawa}},\ }\bibfield  {title} {\bibinfo {title} {{Subbarrier Fusion
  Reactions and Many-Particle Quantum Tunneling}},\ }\href
  {https://doi.org/10.1143/PTP.128.1061} {\bibfield  {journal} {\bibinfo
  {journal} {Progress of Theoretical Physics}\ }\textbf {\bibinfo {volume}
  {128}},\ \bibinfo {pages} {1061} (\bibinfo {year} {2012})}\BibitemShut
  {NoStop}%
\bibitem [{\citenamefont {Binnig}\ \emph {et~al.}(1982)\citenamefont {Binnig},
  \citenamefont {Rohrer}, \citenamefont {Gerber},\ and\ \citenamefont
  {Weibel}}]{PhysRevLett.49.57}%
  \BibitemOpen
  \bibfield  {author} {\bibinfo {author} {\bibfnamefont {G.}~\bibnamefont
  {Binnig}}, \bibinfo {author} {\bibfnamefont {H.}~\bibnamefont {Rohrer}},
  \bibinfo {author} {\bibfnamefont {C.}~\bibnamefont {Gerber}},\ and\ \bibinfo
  {author} {\bibfnamefont {E.}~\bibnamefont {Weibel}},\ }\bibfield  {title}
  {\bibinfo {title} {Surface studies by scanning tunneling microscopy},\ }\href
  {https://doi.org/10.1103/PhysRevLett.49.57} {\bibfield  {journal} {\bibinfo
  {journal} {Phys. Rev. Lett.}\ }\textbf {\bibinfo {volume} {49}},\ \bibinfo
  {pages} {57} (\bibinfo {year} {1982})}\BibitemShut {NoStop}%
\bibitem [{\citenamefont {Tersoff}\ and\ \citenamefont
  {Hamann}(1983)}]{PhysRevLett.50.1998}%
  \BibitemOpen
  \bibfield  {author} {\bibinfo {author} {\bibfnamefont {J.}~\bibnamefont
  {Tersoff}}\ and\ \bibinfo {author} {\bibfnamefont {D.~R.}\ \bibnamefont
  {Hamann}},\ }\bibfield  {title} {\bibinfo {title} {Theory and application for
  the scanning tunneling microscope},\ }\href
  {https://doi.org/10.1103/PhysRevLett.50.1998} {\bibfield  {journal} {\bibinfo
   {journal} {Phys. Rev. Lett.}\ }\textbf {\bibinfo {volume} {50}},\ \bibinfo
  {pages} {1998} (\bibinfo {year} {1983})}\BibitemShut {NoStop}%
\bibitem [{\citenamefont {Pommier}\ \emph {et~al.}(2019)\citenamefont
  {Pommier}, \citenamefont {Bretel}, \citenamefont {L\'opez}, \citenamefont
  {Fabre}, \citenamefont {Mayne}, \citenamefont {Boer-Duchemin}, \citenamefont
  {Dujardin}, \citenamefont {Schull}, \citenamefont {Berciaud},\ and\
  \citenamefont {Le~Moal}}]{PhysRevLett.123.027402}%
  \BibitemOpen
  \bibfield  {author} {\bibinfo {author} {\bibfnamefont {D.}~\bibnamefont
  {Pommier}}, \bibinfo {author} {\bibfnamefont {R.}~\bibnamefont {Bretel}},
  \bibinfo {author} {\bibfnamefont {L.~E.~P.}\ \bibnamefont {L\'opez}},
  \bibinfo {author} {\bibfnamefont {F.}~\bibnamefont {Fabre}}, \bibinfo
  {author} {\bibfnamefont {A.}~\bibnamefont {Mayne}}, \bibinfo {author}
  {\bibfnamefont {E.}~\bibnamefont {Boer-Duchemin}}, \bibinfo {author}
  {\bibfnamefont {G.}~\bibnamefont {Dujardin}}, \bibinfo {author}
  {\bibfnamefont {G.}~\bibnamefont {Schull}}, \bibinfo {author} {\bibfnamefont
  {S.}~\bibnamefont {Berciaud}},\ and\ \bibinfo {author} {\bibfnamefont
  {E.}~\bibnamefont {Le~Moal}},\ }\bibfield  {title} {\bibinfo {title}
  {Scanning tunneling microscope-induced excitonic luminescence of a
  two-dimensional semiconductor},\ }\href
  {https://doi.org/10.1103/PhysRevLett.123.027402} {\bibfield  {journal}
  {\bibinfo  {journal} {Phys. Rev. Lett.}\ }\textbf {\bibinfo {volume} {123}},\
  \bibinfo {pages} {027402} (\bibinfo {year} {2019})}\BibitemShut {NoStop}%
\bibitem [{\citenamefont {Strambini}\ \emph {et~al.}(2022)\citenamefont
  {Strambini}, \citenamefont {Spies}, \citenamefont {Ligato}, \citenamefont
  {Ili{\'c}}, \citenamefont {Rouco}, \citenamefont {Gonz{\'a}lez-Orellana},
  \citenamefont {Ilyn}, \citenamefont {Rogero}, \citenamefont {Bergeret},
  \citenamefont {Moodera} \emph {et~al.}}]{strambini2022superconducting}%
  \BibitemOpen
  \bibfield  {author} {\bibinfo {author} {\bibfnamefont {E.}~\bibnamefont
  {Strambini}}, \bibinfo {author} {\bibfnamefont {M.}~\bibnamefont {Spies}},
  \bibinfo {author} {\bibfnamefont {N.}~\bibnamefont {Ligato}}, \bibinfo
  {author} {\bibfnamefont {S.}~\bibnamefont {Ili{\'c}}}, \bibinfo {author}
  {\bibfnamefont {M.}~\bibnamefont {Rouco}}, \bibinfo {author} {\bibfnamefont
  {C.}~\bibnamefont {Gonz{\'a}lez-Orellana}}, \bibinfo {author} {\bibfnamefont
  {M.}~\bibnamefont {Ilyn}}, \bibinfo {author} {\bibfnamefont {C.}~\bibnamefont
  {Rogero}}, \bibinfo {author} {\bibfnamefont {F.}~\bibnamefont {Bergeret}},
  \bibinfo {author} {\bibfnamefont {J.}~\bibnamefont {Moodera}}, \emph
  {et~al.},\ }\bibfield  {title} {\bibinfo {title} {Superconducting spintronic
  tunnel diode},\ }\href {https://doi.org/10.1038/s41467-022-29990-2}
  {\bibfield  {journal} {\bibinfo  {journal} {Nature Communications}\ }\textbf
  {\bibinfo {volume} {13}},\ \bibinfo {pages} {2431} (\bibinfo {year}
  {2022})}\BibitemShut {NoStop}%
\bibitem [{\citenamefont {Bouscher}\ \emph {et~al.}(2022)\citenamefont
  {Bouscher}, \citenamefont {Panna}, \citenamefont {Balasubramanian},
  \citenamefont {Cohen}, \citenamefont {Ritter},\ and\ \citenamefont
  {Hayat}}]{PhysRevLett.128.127701}%
  \BibitemOpen
  \bibfield  {author} {\bibinfo {author} {\bibfnamefont {S.}~\bibnamefont
  {Bouscher}}, \bibinfo {author} {\bibfnamefont {D.}~\bibnamefont {Panna}},
  \bibinfo {author} {\bibfnamefont {K.}~\bibnamefont {Balasubramanian}},
  \bibinfo {author} {\bibfnamefont {S.}~\bibnamefont {Cohen}}, \bibinfo
  {author} {\bibfnamefont {D.}~\bibnamefont {Ritter}},\ and\ \bibinfo {author}
  {\bibfnamefont {A.}~\bibnamefont {Hayat}},\ }\bibfield  {title} {\bibinfo
  {title} {Enhanced cooper-pair injection into a semiconductor structure by
  resonant tunneling},\ }\href {https://doi.org/10.1103/PhysRevLett.128.127701}
  {\bibfield  {journal} {\bibinfo  {journal} {Phys. Rev. Lett.}\ }\textbf
  {\bibinfo {volume} {128}},\ \bibinfo {pages} {127701} (\bibinfo {year}
  {2022})}\BibitemShut {NoStop}%
\bibitem [{\citenamefont {L\"owdin}(1963)}]{RevModPhys.35.724}%
  \BibitemOpen
  \bibfield  {author} {\bibinfo {author} {\bibfnamefont {P.-O.}\ \bibnamefont
  {L\"owdin}},\ }\bibfield  {title} {\bibinfo {title} {Proton tunneling in
  {DNA} and its biological implications},\ }\href
  {https://doi.org/10.1103/RevModPhys.35.724} {\bibfield  {journal} {\bibinfo
  {journal} {Rev. Mod. Phys.}\ }\textbf {\bibinfo {volume} {35}},\ \bibinfo
  {pages} {724} (\bibinfo {year} {1963})}\BibitemShut {NoStop}%
\bibitem [{\citenamefont {Cao}\ \emph {et~al.}(2020)\citenamefont {Cao},
  \citenamefont {Cogdell}, \citenamefont {Coker}, \citenamefont {Duan},
  \citenamefont {Hauer}, \citenamefont {Kleinekathöfer}, \citenamefont
  {Jansen}, \citenamefont {Mančal}, \citenamefont {Miller}, \citenamefont
  {Ogilvie}, \citenamefont {Prokhorenko}, \citenamefont {Renger}, \citenamefont
  {Tan}, \citenamefont {Tempelaar}, \citenamefont {Thorwart}, \citenamefont
  {Thyrhaug}, \citenamefont {Westenhoff},\ and\ \citenamefont
  {Zigmantas}}]{doi:10.1126/sciadv.aaz4888}%
  \BibitemOpen
  \bibfield  {author} {\bibinfo {author} {\bibfnamefont {J.}~\bibnamefont
  {Cao}}, \bibinfo {author} {\bibfnamefont {R.~J.}\ \bibnamefont {Cogdell}},
  \bibinfo {author} {\bibfnamefont {D.~F.}\ \bibnamefont {Coker}}, \bibinfo
  {author} {\bibfnamefont {H.-G.}\ \bibnamefont {Duan}}, \bibinfo {author}
  {\bibfnamefont {J.}~\bibnamefont {Hauer}}, \bibinfo {author} {\bibfnamefont
  {U.}~\bibnamefont {Kleinekathöfer}}, \bibinfo {author} {\bibfnamefont
  {T.~L.~C.}\ \bibnamefont {Jansen}}, \bibinfo {author} {\bibfnamefont
  {T.}~\bibnamefont {Mančal}}, \bibinfo {author} {\bibfnamefont {R.~J.~D.}\
  \bibnamefont {Miller}}, \bibinfo {author} {\bibfnamefont {J.~P.}\
  \bibnamefont {Ogilvie}}, \bibinfo {author} {\bibfnamefont {V.~I.}\
  \bibnamefont {Prokhorenko}}, \bibinfo {author} {\bibfnamefont
  {T.}~\bibnamefont {Renger}}, \bibinfo {author} {\bibfnamefont {H.-S.}\
  \bibnamefont {Tan}}, \bibinfo {author} {\bibfnamefont {R.}~\bibnamefont
  {Tempelaar}}, \bibinfo {author} {\bibfnamefont {M.}~\bibnamefont {Thorwart}},
  \bibinfo {author} {\bibfnamefont {E.}~\bibnamefont {Thyrhaug}}, \bibinfo
  {author} {\bibfnamefont {S.}~\bibnamefont {Westenhoff}},\ and\ \bibinfo
  {author} {\bibfnamefont {D.}~\bibnamefont {Zigmantas}},\ }\bibfield  {title}
  {\bibinfo {title} {Quantum biology revisited},\ }\href
  {https://doi.org/10.1126/sciadv.aaz4888} {\bibfield  {journal} {\bibinfo
  {journal} {Science Advances}\ }\textbf {\bibinfo {volume} {6}},\ \bibinfo
  {pages} {eaaz4888} (\bibinfo {year} {2020})}\BibitemShut {NoStop}%
\bibitem [{\citenamefont {Kim}\ \emph {et~al.}(2021)\citenamefont {Kim},
  \citenamefont {Bertagna}, \citenamefont {D’Souza}, \citenamefont {Heyes},
  \citenamefont {Johannissen}, \citenamefont {Nery}, \citenamefont {Pantelias},
  \citenamefont {Sanchez-Pedreño~Jimenez}, \citenamefont {Slocombe},
  \citenamefont {Spencer}, \citenamefont {Al-Khalili}, \citenamefont {Engel},
  \citenamefont {Hay}, \citenamefont {Hingley-Wilson}, \citenamefont
  {Jeevaratnam}, \citenamefont {Jones}, \citenamefont {Kattnig}, \citenamefont
  {Lewis}, \citenamefont {Sacchi}, \citenamefont {Scrutton}, \citenamefont
  {Silva},\ and\ \citenamefont {McFadden}}]{quantum3010006}%
  \BibitemOpen
  \bibfield  {author} {\bibinfo {author} {\bibfnamefont {Y.}~\bibnamefont
  {Kim}}, \bibinfo {author} {\bibfnamefont {F.}~\bibnamefont {Bertagna}},
  \bibinfo {author} {\bibfnamefont {E.~M.}\ \bibnamefont {D’Souza}}, \bibinfo
  {author} {\bibfnamefont {D.~J.}\ \bibnamefont {Heyes}}, \bibinfo {author}
  {\bibfnamefont {L.~O.}\ \bibnamefont {Johannissen}}, \bibinfo {author}
  {\bibfnamefont {E.~T.}\ \bibnamefont {Nery}}, \bibinfo {author}
  {\bibfnamefont {A.}~\bibnamefont {Pantelias}}, \bibinfo {author}
  {\bibfnamefont {A.}~\bibnamefont {Sanchez-Pedreño~Jimenez}}, \bibinfo
  {author} {\bibfnamefont {L.}~\bibnamefont {Slocombe}}, \bibinfo {author}
  {\bibfnamefont {M.~G.}\ \bibnamefont {Spencer}}, \bibinfo {author}
  {\bibfnamefont {J.}~\bibnamefont {Al-Khalili}}, \bibinfo {author}
  {\bibfnamefont {G.~S.}\ \bibnamefont {Engel}}, \bibinfo {author}
  {\bibfnamefont {S.}~\bibnamefont {Hay}}, \bibinfo {author} {\bibfnamefont
  {S.~M.}\ \bibnamefont {Hingley-Wilson}}, \bibinfo {author} {\bibfnamefont
  {K.}~\bibnamefont {Jeevaratnam}}, \bibinfo {author} {\bibfnamefont {A.~R.}\
  \bibnamefont {Jones}}, \bibinfo {author} {\bibfnamefont {D.~R.}\ \bibnamefont
  {Kattnig}}, \bibinfo {author} {\bibfnamefont {R.}~\bibnamefont {Lewis}},
  \bibinfo {author} {\bibfnamefont {M.}~\bibnamefont {Sacchi}}, \bibinfo
  {author} {\bibfnamefont {N.~S.}\ \bibnamefont {Scrutton}}, \bibinfo {author}
  {\bibfnamefont {S.~R.~P.}\ \bibnamefont {Silva}},\ and\ \bibinfo {author}
  {\bibfnamefont {J.}~\bibnamefont {McFadden}},\ }\bibfield  {title} {\bibinfo
  {title} {Quantum biology: An update and perspective},\ }\href
  {https://doi.org/10.3390/quantum3010006} {\bibfield  {journal} {\bibinfo
  {journal} {Quantum Reports}\ }\textbf {\bibinfo {volume} {3}},\ \bibinfo
  {pages} {80} (\bibinfo {year} {2021})}\BibitemShut {NoStop}%
\bibitem [{\citenamefont {Zuev}\ \emph {et~al.}(2003)\citenamefont {Zuev},
  \citenamefont {Sheridan}, \citenamefont {Albu}, \citenamefont {Truhlar},
  \citenamefont {Hrovat},\ and\ \citenamefont
  {Borden}}]{doi:10.1126/science.1079294}%
  \BibitemOpen
  \bibfield  {author} {\bibinfo {author} {\bibfnamefont {P.~S.}\ \bibnamefont
  {Zuev}}, \bibinfo {author} {\bibfnamefont {R.~S.}\ \bibnamefont {Sheridan}},
  \bibinfo {author} {\bibfnamefont {T.~V.}\ \bibnamefont {Albu}}, \bibinfo
  {author} {\bibfnamefont {D.~G.}\ \bibnamefont {Truhlar}}, \bibinfo {author}
  {\bibfnamefont {D.~A.}\ \bibnamefont {Hrovat}},\ and\ \bibinfo {author}
  {\bibfnamefont {W.~T.}\ \bibnamefont {Borden}},\ }\bibfield  {title}
  {\bibinfo {title} {Carbon tunneling from a single quantum state},\ }\href
  {https://doi.org/10.1126/science.1079294} {\bibfield  {journal} {\bibinfo
  {journal} {Science}\ }\textbf {\bibinfo {volume} {299}},\ \bibinfo {pages}
  {867} (\bibinfo {year} {2003})}\BibitemShut {NoStop}%
\bibitem [{\citenamefont {Wild}\ \emph {et~al.}(2023)\citenamefont {Wild},
  \citenamefont {N{\"o}tzold}, \citenamefont {Simpson}, \citenamefont {Tran},\
  and\ \citenamefont {Wester}}]{wild2023tunnelling}%
  \BibitemOpen
  \bibfield  {author} {\bibinfo {author} {\bibfnamefont {R.}~\bibnamefont
  {Wild}}, \bibinfo {author} {\bibfnamefont {M.}~\bibnamefont {N{\"o}tzold}},
  \bibinfo {author} {\bibfnamefont {M.}~\bibnamefont {Simpson}}, \bibinfo
  {author} {\bibfnamefont {T.~D.}\ \bibnamefont {Tran}},\ and\ \bibinfo
  {author} {\bibfnamefont {R.}~\bibnamefont {Wester}},\ }\bibfield  {title}
  {\bibinfo {title} {Tunnelling measured in a very slow ion--molecule
  reaction},\ }\href {https://doi.org/10.1038/s41586-023-05727-z} {\bibfield
  {journal} {\bibinfo  {journal} {Nature}\ }\textbf {\bibinfo {volume} {615}},\
  \bibinfo {pages} {425} (\bibinfo {year} {2023})}\BibitemShut {NoStop}%
\bibitem [{\citenamefont {Suh}\ \emph {et~al.}(2020)\citenamefont {Suh},
  \citenamefont {Faseela}, \citenamefont {Kim}, \citenamefont {Park},
  \citenamefont {Lim}, \citenamefont {Seo}, \citenamefont {Kim}, \citenamefont
  {Moon},\ and\ \citenamefont {Baik}}]{suh2020electron}%
  \BibitemOpen
  \bibfield  {author} {\bibinfo {author} {\bibfnamefont {D.}~\bibnamefont
  {Suh}}, \bibinfo {author} {\bibfnamefont {K.}~\bibnamefont {Faseela}},
  \bibinfo {author} {\bibfnamefont {W.}~\bibnamefont {Kim}}, \bibinfo {author}
  {\bibfnamefont {C.}~\bibnamefont {Park}}, \bibinfo {author} {\bibfnamefont
  {J.~G.}\ \bibnamefont {Lim}}, \bibinfo {author} {\bibfnamefont
  {S.}~\bibnamefont {Seo}}, \bibinfo {author} {\bibfnamefont {M.~K.}\
  \bibnamefont {Kim}}, \bibinfo {author} {\bibfnamefont {H.}~\bibnamefont
  {Moon}},\ and\ \bibinfo {author} {\bibfnamefont {S.}~\bibnamefont {Baik}},\
  }\bibfield  {title} {\bibinfo {title} {Electron tunneling of hierarchically
  structured silver nanosatellite particles for highly conductive healable
  nanocomposites},\ }\href {https://doi.org/10.1038/s41467-020-15709-8}
  {\bibfield  {journal} {\bibinfo  {journal} {Nature Communications}\ }\textbf
  {\bibinfo {volume} {11}},\ \bibinfo {pages} {2252} (\bibinfo {year}
  {2020})}\BibitemShut {NoStop}%
\bibitem [{\citenamefont {Drechsel-Grau}\ and\ \citenamefont
  {Marx}(2014)}]{PhysRevLett.112.148302}%
  \BibitemOpen
  \bibfield  {author} {\bibinfo {author} {\bibfnamefont {C.}~\bibnamefont
  {Drechsel-Grau}}\ and\ \bibinfo {author} {\bibfnamefont {D.}~\bibnamefont
  {Marx}},\ }\bibfield  {title} {\bibinfo {title} {Quantum simulation of
  collective proton tunneling in hexagonal ice crystals},\ }\href
  {https://doi.org/10.1103/PhysRevLett.112.148302} {\bibfield  {journal}
  {\bibinfo  {journal} {Phys. Rev. Lett.}\ }\textbf {\bibinfo {volume} {112}},\
  \bibinfo {pages} {148302} (\bibinfo {year} {2014})}\BibitemShut {NoStop}%
\bibitem [{\citenamefont {Borges}\ \emph {et~al.}(2012)\citenamefont {Borges},
  \citenamefont {Sanz}, \citenamefont {Villas-B\^oas}, \citenamefont
  {Diniz~Neto},\ and\ \citenamefont {Alcalde}}]{PhysRevB.85.115425}%
  \BibitemOpen
  \bibfield  {author} {\bibinfo {author} {\bibfnamefont {H.~S.}\ \bibnamefont
  {Borges}}, \bibinfo {author} {\bibfnamefont {L.}~\bibnamefont {Sanz}},
  \bibinfo {author} {\bibfnamefont {J.~M.}\ \bibnamefont {Villas-B\^oas}},
  \bibinfo {author} {\bibfnamefont {O.~O.}\ \bibnamefont {Diniz~Neto}},\ and\
  \bibinfo {author} {\bibfnamefont {A.~M.}\ \bibnamefont {Alcalde}},\
  }\bibfield  {title} {\bibinfo {title} {Tunneling induced transparency and
  slow light in quantum dot molecules},\ }\href
  {https://doi.org/10.1103/PhysRevB.85.115425} {\bibfield  {journal} {\bibinfo
  {journal} {Phys. Rev. B}\ }\textbf {\bibinfo {volume} {85}},\ \bibinfo
  {pages} {115425} (\bibinfo {year} {2012})}\BibitemShut {NoStop}%
\bibitem [{\citenamefont {Hamdad}\ \emph {et~al.}(2023)\citenamefont {Hamdad},
  \citenamefont {Malchow}, \citenamefont {Avetisyan}, \citenamefont {Dujardin},
  \citenamefont {Bouhelier}, \citenamefont {Zhou}, \citenamefont {Cheng},
  \citenamefont {Zellweger},\ and\ \citenamefont
  {Leuthold}}]{PhysRevApplied.20.024057}%
  \BibitemOpen
  \bibfield  {author} {\bibinfo {author} {\bibfnamefont {S.}~\bibnamefont
  {Hamdad}}, \bibinfo {author} {\bibfnamefont {K.}~\bibnamefont {Malchow}},
  \bibinfo {author} {\bibfnamefont {D.}~\bibnamefont {Avetisyan}}, \bibinfo
  {author} {\bibfnamefont {E.}~\bibnamefont {Dujardin}}, \bibinfo {author}
  {\bibfnamefont {A.}~\bibnamefont {Bouhelier}}, \bibinfo {author}
  {\bibfnamefont {Y.}~\bibnamefont {Zhou}}, \bibinfo {author} {\bibfnamefont
  {B.}~\bibnamefont {Cheng}}, \bibinfo {author} {\bibfnamefont
  {T.}~\bibnamefont {Zellweger}},\ and\ \bibinfo {author} {\bibfnamefont
  {J.}~\bibnamefont {Leuthold}},\ }\bibfield  {title} {\bibinfo {title}
  {Overbias and quantum tunneling in light-emitting memristors},\ }\href
  {https://doi.org/10.1103/PhysRevApplied.20.024057} {\bibfield  {journal}
  {\bibinfo  {journal} {Phys. Rev. Appl.}\ }\textbf {\bibinfo {volume} {20}},\
  \bibinfo {pages} {024057} (\bibinfo {year} {2023})}\BibitemShut {NoStop}%
\bibitem [{\citenamefont {Kim}\ \emph {et~al.}(2004)\citenamefont {Kim},
  \citenamefont {Vishveshwara},\ and\ \citenamefont
  {Fradkin}}]{PhysRevLett.93.266803}%
  \BibitemOpen
  \bibfield  {author} {\bibinfo {author} {\bibfnamefont {E.-A.}\ \bibnamefont
  {Kim}}, \bibinfo {author} {\bibfnamefont {S.}~\bibnamefont {Vishveshwara}},\
  and\ \bibinfo {author} {\bibfnamefont {E.}~\bibnamefont {Fradkin}},\
  }\bibfield  {title} {\bibinfo {title} {Cooper-pair tunneling in junctions of
  singlet quantum {H}all states and superconductors},\ }\href
  {https://doi.org/10.1103/PhysRevLett.93.266803} {\bibfield  {journal}
  {\bibinfo  {journal} {Phys. Rev. Lett.}\ }\textbf {\bibinfo {volume} {93}},\
  \bibinfo {pages} {266803} (\bibinfo {year} {2004})}\BibitemShut {NoStop}%
\bibitem [{\citenamefont {Reiter}\ \emph {et~al.}(2002)\citenamefont {Reiter},
  \citenamefont {Mayers},\ and\ \citenamefont
  {Platzman}}]{PhysRevLett.89.135505}%
  \BibitemOpen
  \bibfield  {author} {\bibinfo {author} {\bibfnamefont {G.~F.}\ \bibnamefont
  {Reiter}}, \bibinfo {author} {\bibfnamefont {J.}~\bibnamefont {Mayers}},\
  and\ \bibinfo {author} {\bibfnamefont {P.}~\bibnamefont {Platzman}},\
  }\bibfield  {title} {\bibinfo {title} {Direct observation of tunneling in
  {KDP} using neutron {C}ompton scattering},\ }\href
  {https://doi.org/10.1103/PhysRevLett.89.135505} {\bibfield  {journal}
  {\bibinfo  {journal} {Phys. Rev. Lett.}\ }\textbf {\bibinfo {volume} {89}},\
  \bibinfo {pages} {135505} (\bibinfo {year} {2002})}\BibitemShut {NoStop}%
\bibitem [{\citenamefont {Della~Valle}\ \emph {et~al.}(2007)\citenamefont
  {Della~Valle}, \citenamefont {Ornigotti}, \citenamefont {Cianci},
  \citenamefont {Foglietti}, \citenamefont {Laporta},\ and\ \citenamefont
  {Longhi}}]{PhysRevLett.98.263601}%
  \BibitemOpen
  \bibfield  {author} {\bibinfo {author} {\bibfnamefont {G.}~\bibnamefont
  {Della~Valle}}, \bibinfo {author} {\bibfnamefont {M.}~\bibnamefont
  {Ornigotti}}, \bibinfo {author} {\bibfnamefont {E.}~\bibnamefont {Cianci}},
  \bibinfo {author} {\bibfnamefont {V.}~\bibnamefont {Foglietti}}, \bibinfo
  {author} {\bibfnamefont {P.}~\bibnamefont {Laporta}},\ and\ \bibinfo {author}
  {\bibfnamefont {S.}~\bibnamefont {Longhi}},\ }\bibfield  {title} {\bibinfo
  {title} {Visualization of coherent destruction of tunneling in an optical
  double well system},\ }\href {https://doi.org/10.1103/PhysRevLett.98.263601}
  {\bibfield  {journal} {\bibinfo  {journal} {Phys. Rev. Lett.}\ }\textbf
  {\bibinfo {volume} {98}},\ \bibinfo {pages} {263601} (\bibinfo {year}
  {2007})}\BibitemShut {NoStop}%
\bibitem [{\citenamefont {Lignier}\ \emph {et~al.}(2007)\citenamefont
  {Lignier}, \citenamefont {Sias}, \citenamefont {Ciampini}, \citenamefont
  {Singh}, \citenamefont {Zenesini}, \citenamefont {Morsch},\ and\
  \citenamefont {Arimondo}}]{PhysRevLett.99.220403}%
  \BibitemOpen
  \bibfield  {author} {\bibinfo {author} {\bibfnamefont {H.}~\bibnamefont
  {Lignier}}, \bibinfo {author} {\bibfnamefont {C.}~\bibnamefont {Sias}},
  \bibinfo {author} {\bibfnamefont {D.}~\bibnamefont {Ciampini}}, \bibinfo
  {author} {\bibfnamefont {Y.}~\bibnamefont {Singh}}, \bibinfo {author}
  {\bibfnamefont {A.}~\bibnamefont {Zenesini}}, \bibinfo {author}
  {\bibfnamefont {O.}~\bibnamefont {Morsch}},\ and\ \bibinfo {author}
  {\bibfnamefont {E.}~\bibnamefont {Arimondo}},\ }\bibfield  {title} {\bibinfo
  {title} {Dynamical control of matter-wave tunneling in periodic potentials},\
  }\href {https://doi.org/10.1103/PhysRevLett.99.220403} {\bibfield  {journal}
  {\bibinfo  {journal} {Phys. Rev. Lett.}\ }\textbf {\bibinfo {volume} {99}},\
  \bibinfo {pages} {220403} (\bibinfo {year} {2007})}\BibitemShut {NoStop}%
\bibitem [{\citenamefont {Kierig}\ \emph {et~al.}(2008)\citenamefont {Kierig},
  \citenamefont {Schnorrberger}, \citenamefont {Schietinger}, \citenamefont
  {Tomkovic},\ and\ \citenamefont {Oberthaler}}]{PhysRevLett.100.190405}%
  \BibitemOpen
  \bibfield  {author} {\bibinfo {author} {\bibfnamefont {E.}~\bibnamefont
  {Kierig}}, \bibinfo {author} {\bibfnamefont {U.}~\bibnamefont
  {Schnorrberger}}, \bibinfo {author} {\bibfnamefont {A.}~\bibnamefont
  {Schietinger}}, \bibinfo {author} {\bibfnamefont {J.}~\bibnamefont
  {Tomkovic}},\ and\ \bibinfo {author} {\bibfnamefont {M.~K.}\ \bibnamefont
  {Oberthaler}},\ }\bibfield  {title} {\bibinfo {title} {Single-particle
  tunneling in strongly driven double-well potentials},\ }\href
  {https://doi.org/10.1103/PhysRevLett.100.190405} {\bibfield  {journal}
  {\bibinfo  {journal} {Phys. Rev. Lett.}\ }\textbf {\bibinfo {volume} {100}},\
  \bibinfo {pages} {190405} (\bibinfo {year} {2008})}\BibitemShut {NoStop}%
\bibitem [{\citenamefont {J\"urgensen}\ \emph {et~al.}(2014)\citenamefont
  {J\"urgensen}, \citenamefont {Meinert}, \citenamefont {Mark}, \citenamefont
  {N\"agerl},\ and\ \citenamefont {L\"uhmann}}]{PhysRevLett.113.193003}%
  \BibitemOpen
  \bibfield  {author} {\bibinfo {author} {\bibfnamefont {O.}~\bibnamefont
  {J\"urgensen}}, \bibinfo {author} {\bibfnamefont {F.}~\bibnamefont
  {Meinert}}, \bibinfo {author} {\bibfnamefont {M.~J.}\ \bibnamefont {Mark}},
  \bibinfo {author} {\bibfnamefont {H.-C.}\ \bibnamefont {N\"agerl}},\ and\
  \bibinfo {author} {\bibfnamefont {D.-S.}\ \bibnamefont {L\"uhmann}},\
  }\bibfield  {title} {\bibinfo {title} {Observation of density-induced
  tunneling},\ }\href {https://doi.org/10.1103/PhysRevLett.113.193003}
  {\bibfield  {journal} {\bibinfo  {journal} {Phys. Rev. Lett.}\ }\textbf
  {\bibinfo {volume} {113}},\ \bibinfo {pages} {193003} (\bibinfo {year}
  {2014})}\BibitemShut {NoStop}%
\bibitem [{\citenamefont {Liu}\ \emph {et~al.}(2022)\citenamefont {Liu},
  \citenamefont {Zhang}, \citenamefont {Wu}, \citenamefont {Wang},\ and\
  \citenamefont {Chen}}]{PhysRevResearch.4.L012043}%
  \BibitemOpen
  \bibfield  {author} {\bibinfo {author} {\bibfnamefont {W.}~\bibnamefont
  {Liu}}, \bibinfo {author} {\bibfnamefont {B.}~\bibnamefont {Zhang}}, \bibinfo
  {author} {\bibfnamefont {B.}~\bibnamefont {Wu}}, \bibinfo {author}
  {\bibfnamefont {L.}~\bibnamefont {Wang}},\ and\ \bibinfo {author}
  {\bibfnamefont {F.}~\bibnamefont {Chen}},\ }\bibfield  {title} {\bibinfo
  {title} {Observation of optical tunneling inhibition by a parabolic potential
  in twisted photonic lattices},\ }\href
  {https://doi.org/10.1103/PhysRevResearch.4.L012043} {\bibfield  {journal}
  {\bibinfo  {journal} {Phys. Rev. Research}\ }\textbf {\bibinfo {volume}
  {4}},\ \bibinfo {pages} {L012043} (\bibinfo {year} {2022})}\BibitemShut
  {NoStop}%
\bibitem [{\citenamefont {Fortun}\ \emph {et~al.}(2016)\citenamefont {Fortun},
  \citenamefont {Cabrera-Guti\'errez}, \citenamefont {Condon}, \citenamefont
  {Michon}, \citenamefont {Billy},\ and\ \citenamefont
  {Gu\'ery-Odelin}}]{PhysRevLett.117.010401}%
  \BibitemOpen
  \bibfield  {author} {\bibinfo {author} {\bibfnamefont {A.}~\bibnamefont
  {Fortun}}, \bibinfo {author} {\bibfnamefont {C.}~\bibnamefont
  {Cabrera-Guti\'errez}}, \bibinfo {author} {\bibfnamefont {G.}~\bibnamefont
  {Condon}}, \bibinfo {author} {\bibfnamefont {E.}~\bibnamefont {Michon}},
  \bibinfo {author} {\bibfnamefont {J.}~\bibnamefont {Billy}},\ and\ \bibinfo
  {author} {\bibfnamefont {D.}~\bibnamefont {Gu\'ery-Odelin}},\ }\bibfield
  {title} {\bibinfo {title} {Direct tunneling delay time measurement in an
  optical lattice},\ }\href {https://doi.org/10.1103/PhysRevLett.117.010401}
  {\bibfield  {journal} {\bibinfo  {journal} {Phys. Rev. Lett.}\ }\textbf
  {\bibinfo {volume} {117}},\ \bibinfo {pages} {010401} (\bibinfo {year}
  {2016})}\BibitemShut {NoStop}%
\bibitem [{\citenamefont {Fölling}\ \emph {et~al.}(2007)\citenamefont
  {Fölling}, \citenamefont {Trotzky}, \citenamefont {Cheinet}, \citenamefont
  {Feld}, \citenamefont {Saers}, \citenamefont {Widera}, \citenamefont
  {Müller},\ and\ \citenamefont {Bloch}}]{Foelling2007}%
  \BibitemOpen
  \bibfield  {author} {\bibinfo {author} {\bibfnamefont {S.}~\bibnamefont
  {Fölling}}, \bibinfo {author} {\bibfnamefont {S.}~\bibnamefont {Trotzky}},
  \bibinfo {author} {\bibfnamefont {P.}~\bibnamefont {Cheinet}}, \bibinfo
  {author} {\bibfnamefont {M.}~\bibnamefont {Feld}}, \bibinfo {author}
  {\bibfnamefont {R.}~\bibnamefont {Saers}}, \bibinfo {author} {\bibfnamefont
  {A.}~\bibnamefont {Widera}}, \bibinfo {author} {\bibfnamefont
  {T.}~\bibnamefont {Müller}},\ and\ \bibinfo {author} {\bibfnamefont
  {I.}~\bibnamefont {Bloch}},\ }\bibfield  {title} {\bibinfo {title} {Direct
  observation of second-order atom tunnelling},\ }\href
  {https://doi.org/10.1038/nature06112} {\bibfield  {journal} {\bibinfo
  {journal} {Nature}\ }\textbf {\bibinfo {volume} {448}},\ \bibinfo {pages}
  {1029} (\bibinfo {year} {2007})}\BibitemShut {NoStop}%
\bibitem [{\citenamefont {Eckle}\ \emph {et~al.}(2008)\citenamefont {Eckle},
  \citenamefont {Pfeiffer}, \citenamefont {Cirelli}, \citenamefont {Staudte},
  \citenamefont {Dörner}, \citenamefont {Muller}, \citenamefont {Büttiker},\
  and\ \citenamefont {Keller}}]{doi:10.1126/science.1163439}%
  \BibitemOpen
  \bibfield  {author} {\bibinfo {author} {\bibfnamefont {P.}~\bibnamefont
  {Eckle}}, \bibinfo {author} {\bibfnamefont {A.~N.}\ \bibnamefont {Pfeiffer}},
  \bibinfo {author} {\bibfnamefont {C.}~\bibnamefont {Cirelli}}, \bibinfo
  {author} {\bibfnamefont {A.}~\bibnamefont {Staudte}}, \bibinfo {author}
  {\bibfnamefont {R.}~\bibnamefont {Dörner}}, \bibinfo {author} {\bibfnamefont
  {H.~G.}\ \bibnamefont {Muller}}, \bibinfo {author} {\bibfnamefont
  {M.}~\bibnamefont {Büttiker}},\ and\ \bibinfo {author} {\bibfnamefont
  {U.}~\bibnamefont {Keller}},\ }\bibfield  {title} {\bibinfo {title}
  {Attosecond ionization and tunneling delay time measurements in helium},\
  }\href {https://doi.org/10.1126/science.1163439} {\bibfield  {journal}
  {\bibinfo  {journal} {Science}\ }\textbf {\bibinfo {volume} {322}},\ \bibinfo
  {pages} {1525} (\bibinfo {year} {2008})}\BibitemShut {NoStop}%
\bibitem [{\citenamefont {Sainadh}\ \emph {et~al.}(2019)\citenamefont
  {Sainadh}, \citenamefont {Xu}, \citenamefont {Wang}, \citenamefont
  {Atia-Tul-Noor}, \citenamefont {Wallace}, \citenamefont {Douguet},
  \citenamefont {Bray}, \citenamefont {Ivanov}, \citenamefont {Bartschat},
  \citenamefont {Kheifets}, \citenamefont {Sang},\ and\ \citenamefont
  {Litvinyuk}}]{Sainadh2019}%
  \BibitemOpen
  \bibfield  {author} {\bibinfo {author} {\bibfnamefont {U.~S.}\ \bibnamefont
  {Sainadh}}, \bibinfo {author} {\bibfnamefont {H.}~\bibnamefont {Xu}},
  \bibinfo {author} {\bibfnamefont {X.}~\bibnamefont {Wang}}, \bibinfo {author}
  {\bibfnamefont {A.}~\bibnamefont {Atia-Tul-Noor}}, \bibinfo {author}
  {\bibfnamefont {W.~C.}\ \bibnamefont {Wallace}}, \bibinfo {author}
  {\bibfnamefont {N.}~\bibnamefont {Douguet}}, \bibinfo {author} {\bibfnamefont
  {A.}~\bibnamefont {Bray}}, \bibinfo {author} {\bibfnamefont {I.}~\bibnamefont
  {Ivanov}}, \bibinfo {author} {\bibfnamefont {K.}~\bibnamefont {Bartschat}},
  \bibinfo {author} {\bibfnamefont {A.}~\bibnamefont {Kheifets}}, \bibinfo
  {author} {\bibfnamefont {R.~T.}\ \bibnamefont {Sang}},\ and\ \bibinfo
  {author} {\bibfnamefont {I.~V.}\ \bibnamefont {Litvinyuk}},\ }\bibfield
  {title} {\bibinfo {title} {Attosecond angular streaking and tunnelling time
  in atomic hydrogen},\ }\href {https://doi.org/10.1038/s41586-019-1028-3}
  {\bibfield  {journal} {\bibinfo  {journal} {Nature}\ }\textbf {\bibinfo
  {volume} {568}},\ \bibinfo {pages} {75} (\bibinfo {year} {2019})}\BibitemShut
  {NoStop}%
\bibitem [{\citenamefont {Camus}\ \emph {et~al.}(2017)\citenamefont {Camus},
  \citenamefont {Yakaboylu}, \citenamefont {Fechner}, \citenamefont {Klaiber},
  \citenamefont {Laux}, \citenamefont {Mi}, \citenamefont {Hatsagortsyan},
  \citenamefont {Pfeifer}, \citenamefont {Keitel},\ and\ \citenamefont
  {Moshammer}}]{PhysRevLett.119.023201}%
  \BibitemOpen
  \bibfield  {author} {\bibinfo {author} {\bibfnamefont {N.}~\bibnamefont
  {Camus}}, \bibinfo {author} {\bibfnamefont {E.}~\bibnamefont {Yakaboylu}},
  \bibinfo {author} {\bibfnamefont {L.}~\bibnamefont {Fechner}}, \bibinfo
  {author} {\bibfnamefont {M.}~\bibnamefont {Klaiber}}, \bibinfo {author}
  {\bibfnamefont {M.}~\bibnamefont {Laux}}, \bibinfo {author} {\bibfnamefont
  {Y.}~\bibnamefont {Mi}}, \bibinfo {author} {\bibfnamefont {K.~Z.}\
  \bibnamefont {Hatsagortsyan}}, \bibinfo {author} {\bibfnamefont
  {T.}~\bibnamefont {Pfeifer}}, \bibinfo {author} {\bibfnamefont {C.~H.}\
  \bibnamefont {Keitel}},\ and\ \bibinfo {author} {\bibfnamefont
  {R.}~\bibnamefont {Moshammer}},\ }\bibfield  {title} {\bibinfo {title}
  {Experimental evidence for quantum tunneling time},\ }\href
  {https://doi.org/10.1103/PhysRevLett.119.023201} {\bibfield  {journal}
  {\bibinfo  {journal} {Phys. Rev. Lett.}\ }\textbf {\bibinfo {volume} {119}},\
  \bibinfo {pages} {023201} (\bibinfo {year} {2017})}\BibitemShut {NoStop}%
\bibitem [{\citenamefont {Ramos}\ \emph {et~al.}(2020)\citenamefont {Ramos},
  \citenamefont {Spierings}, \citenamefont {Racicot},\ and\ \citenamefont
  {Steinberg}}]{Ramos2020}%
  \BibitemOpen
  \bibfield  {author} {\bibinfo {author} {\bibfnamefont {R.}~\bibnamefont
  {Ramos}}, \bibinfo {author} {\bibfnamefont {D.}~\bibnamefont {Spierings}},
  \bibinfo {author} {\bibfnamefont {I.}~\bibnamefont {Racicot}},\ and\ \bibinfo
  {author} {\bibfnamefont {A.~M.}\ \bibnamefont {Steinberg}},\ }\bibfield
  {title} {\bibinfo {title} {Measurement of the time spent by a tunnelling atom
  within the barrier region},\ }\href
  {https://doi.org/10.1038/s41586-020-2490-7} {\bibfield  {journal} {\bibinfo
  {journal} {Nature}\ }\textbf {\bibinfo {volume} {583}},\ \bibinfo {pages}
  {529} (\bibinfo {year} {2020})}\BibitemShut {NoStop}%
\bibitem [{\citenamefont {Yu}\ \emph {et~al.}(2022{\natexlab{a}})\citenamefont
  {Yu}, \citenamefont {Liu}, \citenamefont {Li}, \citenamefont {Yan},
  \citenamefont {Cao}, \citenamefont {Tan}, \citenamefont {Liang},
  \citenamefont {Guo}, \citenamefont {Cao}, \citenamefont {Lan}, \citenamefont
  {Zhang}, \citenamefont {Zhou},\ and\ \citenamefont {Lu}}]{Yu2022}%
  \BibitemOpen
  \bibfield  {author} {\bibinfo {author} {\bibfnamefont {M.}~\bibnamefont
  {Yu}}, \bibinfo {author} {\bibfnamefont {K.}~\bibnamefont {Liu}}, \bibinfo
  {author} {\bibfnamefont {M.}~\bibnamefont {Li}}, \bibinfo {author}
  {\bibfnamefont {J.}~\bibnamefont {Yan}}, \bibinfo {author} {\bibfnamefont
  {C.}~\bibnamefont {Cao}}, \bibinfo {author} {\bibfnamefont {J.}~\bibnamefont
  {Tan}}, \bibinfo {author} {\bibfnamefont {J.}~\bibnamefont {Liang}}, \bibinfo
  {author} {\bibfnamefont {K.}~\bibnamefont {Guo}}, \bibinfo {author}
  {\bibfnamefont {W.}~\bibnamefont {Cao}}, \bibinfo {author} {\bibfnamefont
  {P.}~\bibnamefont {Lan}}, \bibinfo {author} {\bibfnamefont {Q.}~\bibnamefont
  {Zhang}}, \bibinfo {author} {\bibfnamefont {Y.}~\bibnamefont {Zhou}},\ and\
  \bibinfo {author} {\bibfnamefont {P.}~\bibnamefont {Lu}},\ }\bibfield
  {title} {\bibinfo {title} {Full experimental determination of tunneling time
  with attosecond-scale streaking method},\ }\href
  {https://doi.org/10.1038/s41377-022-00911-8} {\bibfield  {journal} {\bibinfo
  {journal} {Light Sci. Appl.}\ }\textbf {\bibinfo {volume} {11}},\ \bibinfo
  {pages} {215} (\bibinfo {year} {2022}{\natexlab{a}})}\BibitemShut {NoStop}%
\bibitem [{\citenamefont {Deng}\ \emph {et~al.}(2020)\citenamefont {Deng},
  \citenamefont {L\"u}, \citenamefont {Ke}, \citenamefont {Guo},\ and\
  \citenamefont {Zhang}}]{PhysRevB.101.085410}%
  \BibitemOpen
  \bibfield  {author} {\bibinfo {author} {\bibfnamefont {Y.-H.}\ \bibnamefont
  {Deng}}, \bibinfo {author} {\bibfnamefont {H.-F.}\ \bibnamefont {L\"u}},
  \bibinfo {author} {\bibfnamefont {S.-S.}\ \bibnamefont {Ke}}, \bibinfo
  {author} {\bibfnamefont {Y.}~\bibnamefont {Guo}},\ and\ \bibinfo {author}
  {\bibfnamefont {H.-W.}\ \bibnamefont {Zhang}},\ }\bibfield  {title} {\bibinfo
  {title} {Quantum tunneling through a rectangular barrier in multi-{W}eyl
  semimetals},\ }\href {https://doi.org/10.1103/PhysRevB.101.085410} {\bibfield
   {journal} {\bibinfo  {journal} {Phys. Rev. B}\ }\textbf {\bibinfo {volume}
  {101}},\ \bibinfo {pages} {085410} (\bibinfo {year} {2020})}\BibitemShut
  {NoStop}%
\bibitem [{\citenamefont {Yu}\ \emph {et~al.}(2022{\natexlab{b}})\citenamefont
  {Yu}, \citenamefont {Ge}, \citenamefont {Wen}, \citenamefont {Du},
  \citenamefont {Zhai}, \citenamefont {Liu}, \citenamefont {Wang},\ and\
  \citenamefont {Qin}}]{yu2022highly}%
  \BibitemOpen
  \bibfield  {author} {\bibinfo {author} {\bibfnamefont {Q.}~\bibnamefont
  {Yu}}, \bibinfo {author} {\bibfnamefont {R.}~\bibnamefont {Ge}}, \bibinfo
  {author} {\bibfnamefont {J.}~\bibnamefont {Wen}}, \bibinfo {author}
  {\bibfnamefont {T.}~\bibnamefont {Du}}, \bibinfo {author} {\bibfnamefont
  {J.}~\bibnamefont {Zhai}}, \bibinfo {author} {\bibfnamefont {S.}~\bibnamefont
  {Liu}}, \bibinfo {author} {\bibfnamefont {L.}~\bibnamefont {Wang}},\ and\
  \bibinfo {author} {\bibfnamefont {Y.}~\bibnamefont {Qin}},\ }\bibfield
  {title} {\bibinfo {title} {Highly sensitive strain sensors based on
  piezotronic tunneling junction},\ }\href
  {https://doi.org/10.1038/s41467-022-28443-0} {\bibfield  {journal} {\bibinfo
  {journal} {Nature Communications}\ }\textbf {\bibinfo {volume} {13}},\
  \bibinfo {pages} {778} (\bibinfo {year} {2022}{\natexlab{b}})}\BibitemShut
  {NoStop}%
\bibitem [{\citenamefont {Ma}\ \emph {et~al.}(2011)\citenamefont {Ma},
  \citenamefont {Tai}, \citenamefont {Preiss}, \citenamefont {Bakr},
  \citenamefont {Simon},\ and\ \citenamefont
  {Greiner}}]{PhysRevLett.107.095301}%
  \BibitemOpen
  \bibfield  {author} {\bibinfo {author} {\bibfnamefont {R.}~\bibnamefont
  {Ma}}, \bibinfo {author} {\bibfnamefont {M.~E.}\ \bibnamefont {Tai}},
  \bibinfo {author} {\bibfnamefont {P.~M.}\ \bibnamefont {Preiss}}, \bibinfo
  {author} {\bibfnamefont {W.~S.}\ \bibnamefont {Bakr}}, \bibinfo {author}
  {\bibfnamefont {J.}~\bibnamefont {Simon}},\ and\ \bibinfo {author}
  {\bibfnamefont {M.}~\bibnamefont {Greiner}},\ }\bibfield  {title} {\bibinfo
  {title} {Photon-assisted tunneling in a biased strongly correlated {B}ose
  gas},\ }\href {https://doi.org/10.1103/PhysRevLett.107.095301} {\bibfield
  {journal} {\bibinfo  {journal} {Phys. Rev. Lett.}\ }\textbf {\bibinfo
  {volume} {107}},\ \bibinfo {pages} {095301} (\bibinfo {year}
  {2011})}\BibitemShut {NoStop}%
\bibitem [{\citenamefont {Gallego-Marcos}\ \emph {et~al.}(2015)\citenamefont
  {Gallego-Marcos}, \citenamefont {Sánchez},\ and\ \citenamefont
  {Platero}}]{GallegoMarcos2015}%
  \BibitemOpen
  \bibfield  {author} {\bibinfo {author} {\bibfnamefont {F.}~\bibnamefont
  {Gallego-Marcos}}, \bibinfo {author} {\bibfnamefont {R.}~\bibnamefont
  {Sánchez}},\ and\ \bibinfo {author} {\bibfnamefont {G.}~\bibnamefont
  {Platero}},\ }\bibfield  {title} {\bibinfo {title} {{Photon assisted
  long-range tunneling}},\ }\href {https://doi.org/10.1063/1.4913834}
  {\bibfield  {journal} {\bibinfo  {journal} {Journal of Applied Physics}\
  }\textbf {\bibinfo {volume} {117}},\ \bibinfo {pages} {112808} (\bibinfo
  {year} {2015})}\BibitemShut {NoStop}%
\bibitem [{\citenamefont {Braakman}\ \emph {et~al.}(2013)\citenamefont
  {Braakman}, \citenamefont {Barthelemy}, \citenamefont {Reichl}, \citenamefont
  {Wegscheider},\ and\ \citenamefont {Vandersypen}}]{Braakman2013}%
  \BibitemOpen
  \bibfield  {author} {\bibinfo {author} {\bibfnamefont {F.~R.}\ \bibnamefont
  {Braakman}}, \bibinfo {author} {\bibfnamefont {P.}~\bibnamefont
  {Barthelemy}}, \bibinfo {author} {\bibfnamefont {C.}~\bibnamefont {Reichl}},
  \bibinfo {author} {\bibfnamefont {W.}~\bibnamefont {Wegscheider}},\ and\
  \bibinfo {author} {\bibfnamefont {L.~M.~K.}\ \bibnamefont {Vandersypen}},\
  }\bibfield  {title} {\bibinfo {title} {Long-distance coherent coupling in a
  quantum dot array},\ }\href {https://doi.org/10.1038/nnano.2013.67}
  {\bibfield  {journal} {\bibinfo  {journal} {Nature Nanotechnology}\ }\textbf
  {\bibinfo {volume} {8}},\ \bibinfo {pages} {432} (\bibinfo {year}
  {2013})}\BibitemShut {NoStop}%
\bibitem [{\citenamefont {Winkler}\ and\ \citenamefont
  {Gray}(2014)}]{Winkler2014}%
  \BibitemOpen
  \bibfield  {author} {\bibinfo {author} {\bibfnamefont {J.~R.}\ \bibnamefont
  {Winkler}}\ and\ \bibinfo {author} {\bibfnamefont {H.~B.}\ \bibnamefont
  {Gray}},\ }\bibfield  {title} {\bibinfo {title} {Long-range electron
  tunneling},\ }\href {https://doi.org/10.1021/ja500215j} {\bibfield  {journal}
  {\bibinfo  {journal} {J. Am. Chem. Soc.}\ }\textbf {\bibinfo {volume}
  {136}},\ \bibinfo {pages} {2930} (\bibinfo {year} {2014})}\BibitemShut
  {NoStop}%
\bibitem [{\citenamefont {Bueno}(2024)}]{D3CS00662J}%
  \BibitemOpen
  \bibfield  {author} {\bibinfo {author} {\bibfnamefont {P.~R.}\ \bibnamefont
  {Bueno}},\ }\bibfield  {title} {\bibinfo {title} {On the fundamentals of
  quantum rate theory and the long-range electron transport in respiratory
  chains},\ }\href {https://doi.org/10.1039/D3CS00662J} {\bibfield  {journal}
  {\bibinfo  {journal} {Chem. Soc. Rev.}\ }\textbf {\bibinfo {volume} {53}},\
  \bibinfo {pages} {5348} (\bibinfo {year} {2024})}\BibitemShut {NoStop}%
\bibitem [{\citenamefont {Meinert}\ \emph {et~al.}(2014)\citenamefont
  {Meinert}, \citenamefont {Mark}, \citenamefont {Kirilov}, \citenamefont
  {Lauber}, \citenamefont {Weinmann}, \citenamefont {Gröbner}, \citenamefont
  {Daley},\ and\ \citenamefont {Nägerl}}]{doi:10.1126/science.1248402}%
  \BibitemOpen
  \bibfield  {author} {\bibinfo {author} {\bibfnamefont {F.}~\bibnamefont
  {Meinert}}, \bibinfo {author} {\bibfnamefont {M.~J.}\ \bibnamefont {Mark}},
  \bibinfo {author} {\bibfnamefont {E.}~\bibnamefont {Kirilov}}, \bibinfo
  {author} {\bibfnamefont {K.}~\bibnamefont {Lauber}}, \bibinfo {author}
  {\bibfnamefont {P.}~\bibnamefont {Weinmann}}, \bibinfo {author}
  {\bibfnamefont {M.}~\bibnamefont {Gröbner}}, \bibinfo {author}
  {\bibfnamefont {A.~J.}\ \bibnamefont {Daley}},\ and\ \bibinfo {author}
  {\bibfnamefont {H.-C.}\ \bibnamefont {Nägerl}},\ }\bibfield  {title}
  {\bibinfo {title} {Observation of many-body dynamics in long-range tunneling
  after a quantum quench},\ }\href {https://doi.org/10.1126/science.1248402}
  {\bibfield  {journal} {\bibinfo  {journal} {Science}\ }\textbf {\bibinfo
  {volume} {344}},\ \bibinfo {pages} {1259} (\bibinfo {year}
  {2014})}\BibitemShut {NoStop}%
\bibitem [{\citenamefont {Kessing}\ \emph {et~al.}(2022)\citenamefont
  {Kessing}, \citenamefont {Yang}, \citenamefont {Manmana},\ and\ \citenamefont
  {Cao}}]{Kessing2022}%
  \BibitemOpen
  \bibfield  {author} {\bibinfo {author} {\bibfnamefont {R.~K.}\ \bibnamefont
  {Kessing}}, \bibinfo {author} {\bibfnamefont {P.-Y.}\ \bibnamefont {Yang}},
  \bibinfo {author} {\bibfnamefont {S.~R.}\ \bibnamefont {Manmana}},\ and\
  \bibinfo {author} {\bibfnamefont {J.}~\bibnamefont {Cao}},\ }\bibfield
  {title} {\bibinfo {title} {Long-range nonequilibrium coherent tunneling
  induced by fractional vibronic resonances},\ }\href
  {https://doi.org/10.1021/acs.jpclett.2c01455} {\bibfield  {journal} {\bibinfo
   {journal} {J. Phys. Chem. Lett.}\ }\textbf {\bibinfo {volume} {13}},\
  \bibinfo {pages} {6831} (\bibinfo {year} {2022})}\BibitemShut {NoStop}%
\bibitem [{\citenamefont {Martinez}\ \emph {et~al.}(2021)\citenamefont
  {Martinez}, \citenamefont {Giraud}, \citenamefont {Ullmo}, \citenamefont
  {Billy}, \citenamefont {Gu\'ery-Odelin}, \citenamefont {Georgeot},\ and\
  \citenamefont {Lemari\'e}}]{PhysRevLett.126.174102}%
  \BibitemOpen
  \bibfield  {author} {\bibinfo {author} {\bibfnamefont {M.}~\bibnamefont
  {Martinez}}, \bibinfo {author} {\bibfnamefont {O.}~\bibnamefont {Giraud}},
  \bibinfo {author} {\bibfnamefont {D.}~\bibnamefont {Ullmo}}, \bibinfo
  {author} {\bibfnamefont {J.}~\bibnamefont {Billy}}, \bibinfo {author}
  {\bibfnamefont {D.}~\bibnamefont {Gu\'ery-Odelin}}, \bibinfo {author}
  {\bibfnamefont {B.}~\bibnamefont {Georgeot}},\ and\ \bibinfo {author}
  {\bibfnamefont {G.}~\bibnamefont {Lemari\'e}},\ }\bibfield  {title} {\bibinfo
  {title} {Chaos-assisted long-range tunneling for quantum simulation},\ }\href
  {https://doi.org/10.1103/PhysRevLett.126.174102} {\bibfield  {journal}
  {\bibinfo  {journal} {Phys. Rev. Lett.}\ }\textbf {\bibinfo {volume} {126}},\
  \bibinfo {pages} {174102} (\bibinfo {year} {2021})}\BibitemShut {NoStop}%
\bibitem [{\citenamefont {Krinner}\ \emph {et~al.}(2018)\citenamefont
  {Krinner}, \citenamefont {Stewart}, \citenamefont {Pazmino}, \citenamefont
  {Kwon},\ and\ \citenamefont {Schneble}}]{krinner2018spontaneous}%
  \BibitemOpen
  \bibfield  {author} {\bibinfo {author} {\bibfnamefont {L.}~\bibnamefont
  {Krinner}}, \bibinfo {author} {\bibfnamefont {M.}~\bibnamefont {Stewart}},
  \bibinfo {author} {\bibfnamefont {A.}~\bibnamefont {Pazmino}}, \bibinfo
  {author} {\bibfnamefont {J.}~\bibnamefont {Kwon}},\ and\ \bibinfo {author}
  {\bibfnamefont {D.}~\bibnamefont {Schneble}},\ }\bibfield  {title} {\bibinfo
  {title} {Spontaneous emission of matter waves from a tunable open quantum
  system},\ }\href {https://doi.org/10.1038/s41586-018-0348-z} {\bibfield
  {journal} {\bibinfo  {journal} {Nature}\ }\textbf {\bibinfo {volume} {559}},\
  \bibinfo {pages} {589} (\bibinfo {year} {2018})}\BibitemShut {NoStop}%
\bibitem [{\citenamefont {Stewart}\ \emph {et~al.}(2020)\citenamefont
  {Stewart}, \citenamefont {Kwon}, \citenamefont {Lanuza},\ and\ \citenamefont
  {Schneble}}]{PhysRevResearch.2.043307}%
  \BibitemOpen
  \bibfield  {author} {\bibinfo {author} {\bibfnamefont {M.}~\bibnamefont
  {Stewart}}, \bibinfo {author} {\bibfnamefont {J.}~\bibnamefont {Kwon}},
  \bibinfo {author} {\bibfnamefont {A.}~\bibnamefont {Lanuza}},\ and\ \bibinfo
  {author} {\bibfnamefont {D.}~\bibnamefont {Schneble}},\ }\bibfield  {title}
  {\bibinfo {title} {Dynamics of matter-wave quantum emitters in a structured
  vacuum},\ }\href {https://doi.org/10.1103/PhysRevResearch.2.043307}
  {\bibfield  {journal} {\bibinfo  {journal} {Phys. Rev. Research}\ }\textbf
  {\bibinfo {volume} {2}},\ \bibinfo {pages} {043307} (\bibinfo {year}
  {2020})}\BibitemShut {NoStop}%
\bibitem [{\citenamefont {Kwon}\ \emph {et~al.}(2022)\citenamefont {Kwon},
  \citenamefont {Kim}, \citenamefont {Lanuza},\ and\ \citenamefont
  {Schneble}}]{kwon2022formation}%
  \BibitemOpen
  \bibfield  {author} {\bibinfo {author} {\bibfnamefont {J.}~\bibnamefont
  {Kwon}}, \bibinfo {author} {\bibfnamefont {Y.}~\bibnamefont {Kim}}, \bibinfo
  {author} {\bibfnamefont {A.}~\bibnamefont {Lanuza}},\ and\ \bibinfo {author}
  {\bibfnamefont {D.}~\bibnamefont {Schneble}},\ }\bibfield  {title} {\bibinfo
  {title} {Formation of matter-wave polaritons in an optical lattice},\ }\href
  {https://doi.org/10.1038/s41567-022-01565-4} {\bibfield  {journal} {\bibinfo
  {journal} {Nature Physics}\ }\textbf {\bibinfo {volume} {18}},\ \bibinfo
  {pages} {657} (\bibinfo {year} {2022})}\BibitemShut {NoStop}%
\bibitem [{\citenamefont {de~Vega}\ \emph {et~al.}(2008)\citenamefont
  {de~Vega}, \citenamefont {Porras},\ and\ \citenamefont
  {Ignacio~Cirac}}]{PhysRevLett.101.260404}%
  \BibitemOpen
  \bibfield  {author} {\bibinfo {author} {\bibfnamefont {I.}~\bibnamefont
  {de~Vega}}, \bibinfo {author} {\bibfnamefont {D.}~\bibnamefont {Porras}},\
  and\ \bibinfo {author} {\bibfnamefont {J.}~\bibnamefont {Ignacio~Cirac}},\
  }\bibfield  {title} {\bibinfo {title} {Matter-wave emission in optical
  lattices: Single particle and collective effects},\ }\href
  {https://doi.org/10.1103/PhysRevLett.101.260404} {\bibfield  {journal}
  {\bibinfo  {journal} {Phys. Rev. Lett.}\ }\textbf {\bibinfo {volume} {101}},\
  \bibinfo {pages} {260404} (\bibinfo {year} {2008})}\BibitemShut {NoStop}%
\bibitem [{\citenamefont {Navarrete-Benlloch}\ \emph
  {et~al.}(2011)\citenamefont {Navarrete-Benlloch}, \citenamefont {de~Vega},
  \citenamefont {Porras},\ and\ \citenamefont
  {Cirac}}]{Navarrete_Benlloch_2011}%
  \BibitemOpen
  \bibfield  {author} {\bibinfo {author} {\bibfnamefont {C.}~\bibnamefont
  {Navarrete-Benlloch}}, \bibinfo {author} {\bibfnamefont {I.}~\bibnamefont
  {de~Vega}}, \bibinfo {author} {\bibfnamefont {D.}~\bibnamefont {Porras}},\
  and\ \bibinfo {author} {\bibfnamefont {J.~I.}\ \bibnamefont {Cirac}},\
  }\bibfield  {title} {\bibinfo {title} {Simulating quantum-optical phenomena
  with cold atoms in optical lattices},\ }\href
  {https://doi.org/10.1088/1367-2630/13/2/023024} {\bibfield  {journal}
  {\bibinfo  {journal} {New Journal of Physics}\ }\textbf {\bibinfo {volume}
  {13}},\ \bibinfo {pages} {023024} (\bibinfo {year} {2011})}\BibitemShut
  {NoStop}%
\bibitem [{\citenamefont {Stewart}\ \emph {et~al.}(2017)\citenamefont
  {Stewart}, \citenamefont {Krinner}, \citenamefont {Pazmi\~no},\ and\
  \citenamefont {Schneble}}]{PhysRevA.95.013626}%
  \BibitemOpen
  \bibfield  {author} {\bibinfo {author} {\bibfnamefont {M.}~\bibnamefont
  {Stewart}}, \bibinfo {author} {\bibfnamefont {L.}~\bibnamefont {Krinner}},
  \bibinfo {author} {\bibfnamefont {A.}~\bibnamefont {Pazmi\~no}},\ and\
  \bibinfo {author} {\bibfnamefont {D.}~\bibnamefont {Schneble}},\ }\bibfield
  {title} {\bibinfo {title} {Analysis of non-{M}arkovian coupling of a
  lattice-trapped atom to free space},\ }\href
  {https://doi.org/10.1103/PhysRevA.95.013626} {\bibfield  {journal} {\bibinfo
  {journal} {Phys. Rev. A}\ }\textbf {\bibinfo {volume} {95}},\ \bibinfo
  {pages} {013626} (\bibinfo {year} {2017})}\BibitemShut {NoStop}%
\bibitem [{\citenamefont {Gonz\'alez-Tudela}\ \emph {et~al.}(2019)\citenamefont
  {Gonz\'alez-Tudela}, \citenamefont {Mu\~noz},\ and\ \citenamefont
  {Cirac}}]{PhysRevLett.122.203603}%
  \BibitemOpen
  \bibfield  {author} {\bibinfo {author} {\bibfnamefont {A.}~\bibnamefont
  {Gonz\'alez-Tudela}}, \bibinfo {author} {\bibfnamefont {C.~S.}\ \bibnamefont
  {Mu\~noz}},\ and\ \bibinfo {author} {\bibfnamefont {J.~I.}\ \bibnamefont
  {Cirac}},\ }\bibfield  {title} {\bibinfo {title} {Engineering and harnessing
  giant atoms in high-dimensional baths: A proposal for implementation with
  cold atoms},\ }\href {https://doi.org/10.1103/PhysRevLett.122.203603}
  {\bibfield  {journal} {\bibinfo  {journal} {Phys. Rev. Lett.}\ }\textbf
  {\bibinfo {volume} {122}},\ \bibinfo {pages} {203603} (\bibinfo {year}
  {2019})}\BibitemShut {NoStop}%
\bibitem [{\citenamefont {Gonz{\'{a}}lez-Tudela}\ and\ \citenamefont
  {Cirac}(2018)}]{GonzalezTudela2018nonmarkovianquantum}%
  \BibitemOpen
  \bibfield  {author} {\bibinfo {author} {\bibfnamefont {A.}~\bibnamefont
  {Gonz{\'{a}}lez-Tudela}}\ and\ \bibinfo {author} {\bibfnamefont {J.~I.}\
  \bibnamefont {Cirac}},\ }\bibfield  {title} {\bibinfo {title}
  {Non-{M}arkovian {Q}uantum {O}ptics with {T}hree-{D}imensional
  {S}tate-{D}ependent {O}ptical {L}attices},\ }\href
  {https://doi.org/10.22331/q-2018-10-01-97} {\bibfield  {journal} {\bibinfo
  {journal} {{Quantum}}\ }\textbf {\bibinfo {volume} {2}},\ \bibinfo {pages}
  {97} (\bibinfo {year} {2018})}\BibitemShut {NoStop}%
\bibitem [{\citenamefont {Bello}\ \emph {et~al.}(2019)\citenamefont {Bello},
  \citenamefont {Platero}, \citenamefont {Cirac},\ and\ \citenamefont
  {Gonz{\'a}lez-Tudela}}]{bello2019unconventional}%
  \BibitemOpen
  \bibfield  {author} {\bibinfo {author} {\bibfnamefont {M.}~\bibnamefont
  {Bello}}, \bibinfo {author} {\bibfnamefont {G.}~\bibnamefont {Platero}},
  \bibinfo {author} {\bibfnamefont {J.~I.}\ \bibnamefont {Cirac}},\ and\
  \bibinfo {author} {\bibfnamefont {A.}~\bibnamefont {Gonz{\'a}lez-Tudela}},\
  }\bibfield  {title} {\bibinfo {title} {Unconventional quantum optics in
  topological waveguide qed},\ }\href
  {https://www.science.org/doi/abs/10.1126/sciadv.aaw0297} {\bibfield
  {journal} {\bibinfo  {journal} {Science Advances}\ }\textbf {\bibinfo
  {volume} {5}},\ \bibinfo {pages} {eaaw0297} (\bibinfo {year}
  {2019})}\BibitemShut {NoStop}%
\bibitem [{\citenamefont {Bloch}\ \emph {et~al.}(2008)\citenamefont {Bloch},
  \citenamefont {Dalibard},\ and\ \citenamefont {Zwerger}}]{RevModPhys.80.885}%
  \BibitemOpen
  \bibfield  {author} {\bibinfo {author} {\bibfnamefont {I.}~\bibnamefont
  {Bloch}}, \bibinfo {author} {\bibfnamefont {J.}~\bibnamefont {Dalibard}},\
  and\ \bibinfo {author} {\bibfnamefont {W.}~\bibnamefont {Zwerger}},\
  }\bibfield  {title} {\bibinfo {title} {Many-body physics with ultracold
  gases},\ }\href {https://doi.org/10.1103/RevModPhys.80.885} {\bibfield
  {journal} {\bibinfo  {journal} {Rev. Mod. Phys.}\ }\textbf {\bibinfo {volume}
  {80}},\ \bibinfo {pages} {885} (\bibinfo {year} {2008})}\BibitemShut
  {NoStop}%
\bibitem [{\citenamefont {Lanuza}\ \emph {et~al.}(2022)\citenamefont {Lanuza},
  \citenamefont {Kwon}, \citenamefont {Kim},\ and\ \citenamefont
  {Schneble}}]{PhysRevA.105.023703}%
  \BibitemOpen
  \bibfield  {author} {\bibinfo {author} {\bibfnamefont {A.}~\bibnamefont
  {Lanuza}}, \bibinfo {author} {\bibfnamefont {J.}~\bibnamefont {Kwon}},
  \bibinfo {author} {\bibfnamefont {Y.}~\bibnamefont {Kim}},\ and\ \bibinfo
  {author} {\bibfnamefont {D.}~\bibnamefont {Schneble}},\ }\bibfield  {title}
  {\bibinfo {title} {Multiband and array effects in matter-wave-based waveguide
  {QED}},\ }\href {https://doi.org/10.1103/PhysRevA.105.023703} {\bibfield
  {journal} {\bibinfo  {journal} {Phys. Rev. A}\ }\textbf {\bibinfo {volume}
  {105}},\ \bibinfo {pages} {023703} (\bibinfo {year} {2022})}\BibitemShut
  {NoStop}%
\bibitem [{\citenamefont {Tavis}\ and\ \citenamefont
  {Cummings}(1968)}]{PhysRev.170.379}%
  \BibitemOpen
  \bibfield  {author} {\bibinfo {author} {\bibfnamefont {M.}~\bibnamefont
  {Tavis}}\ and\ \bibinfo {author} {\bibfnamefont {F.~W.}\ \bibnamefont
  {Cummings}},\ }\bibfield  {title} {\bibinfo {title} {Exact solution for an
  $n$-molecule---radiation-field {H}amiltonian},\ }\href
  {https://doi.org/10.1103/PhysRev.170.379} {\bibfield  {journal} {\bibinfo
  {journal} {Phys. Rev.}\ }\textbf {\bibinfo {volume} {170}},\ \bibinfo {pages}
  {379} (\bibinfo {year} {1968})}\BibitemShut {NoStop}%
\bibitem [{\citenamefont {Davies}(1974)}]{Davies1974}%
  \BibitemOpen
  \bibfield  {author} {\bibinfo {author} {\bibfnamefont {E.~B.}\ \bibnamefont
  {Davies}},\ }\bibfield  {title} {\bibinfo {title} {Markovian master
  equations},\ }\href {https://doi.org/10.1007/BF01608389} {\bibfield
  {journal} {\bibinfo  {journal} {Communications in Mathematical Physics}\
  }\textbf {\bibinfo {volume} {39}},\ \bibinfo {pages} {91} (\bibinfo {year}
  {1974})}\BibitemShut {NoStop}%
\bibitem [{\citenamefont {D\"{u}mcke}\ and\ \citenamefont
  {Spohn}()}]{Duemcke1979}%
  \BibitemOpen
  \bibfield  {author} {\bibinfo {author} {\bibfnamefont {R.}~\bibnamefont
  {D\"{u}mcke}}\ and\ \bibinfo {author} {\bibfnamefont {H.}~\bibnamefont
  {Spohn}},\ }\bibfield  {title} {\bibinfo {title} {The proper form of the
  generator in the weak coupling limit},\ }\href
  {https://doi.org/10.1007/BF01325208} {\bibfield  {journal} {\bibinfo
  {journal} {Zeitschrift f\"{u}r Physik B Condensed Matter}\ }\textbf {\bibinfo
  {volume} {34}},\ \bibinfo {pages} {419}}\BibitemShut {NoStop}%
\bibitem [{\citenamefont {Shahmoon}\ and\ \citenamefont
  {Kurizki}(2013)}]{PhysRevA.87.033831}%
  \BibitemOpen
  \bibfield  {author} {\bibinfo {author} {\bibfnamefont {E.}~\bibnamefont
  {Shahmoon}}\ and\ \bibinfo {author} {\bibfnamefont {G.}~\bibnamefont
  {Kurizki}},\ }\bibfield  {title} {\bibinfo {title} {Nonradiative interaction
  and entanglement between distant atoms},\ }\href
  {https://doi.org/10.1103/PhysRevA.87.033831} {\bibfield  {journal} {\bibinfo
  {journal} {Phys. Rev. A}\ }\textbf {\bibinfo {volume} {87}},\ \bibinfo
  {pages} {033831} (\bibinfo {year} {2013})}\BibitemShut {NoStop}%
\bibitem [{\citenamefont {Yang}\ \emph {et~al.}(2019)\citenamefont {Yang},
  \citenamefont {An},\ and\ \citenamefont {Lin}}]{PhysRevResearch.1.023027}%
  \BibitemOpen
  \bibfield  {author} {\bibinfo {author} {\bibfnamefont {C.-J.}\ \bibnamefont
  {Yang}}, \bibinfo {author} {\bibfnamefont {J.-H.}\ \bibnamefont {An}},\ and\
  \bibinfo {author} {\bibfnamefont {H.-Q.}\ \bibnamefont {Lin}},\ }\bibfield
  {title} {\bibinfo {title} {Signatures of quantized coupling between quantum
  emitters and localized surface plasmons},\ }\href
  {https://doi.org/10.1103/PhysRevResearch.1.023027} {\bibfield  {journal}
  {\bibinfo  {journal} {Phys. Rev. Research}\ }\textbf {\bibinfo {volume}
  {1}},\ \bibinfo {pages} {023027} (\bibinfo {year} {2019})}\BibitemShut
  {NoStop}%
\bibitem [{\citenamefont {Wu}\ and\ \citenamefont
  {An}(2021)}]{PhysRevA.104.042609}%
  \BibitemOpen
  \bibfield  {author} {\bibinfo {author} {\bibfnamefont {W.}~\bibnamefont
  {Wu}}\ and\ \bibinfo {author} {\bibfnamefont {J.-H.}\ \bibnamefont {An}},\
  }\bibfield  {title} {\bibinfo {title} {Gaussian quantum metrology in a
  dissipative environment},\ }\href
  {https://doi.org/10.1103/PhysRevA.104.042609} {\bibfield  {journal} {\bibinfo
   {journal} {Phys. Rev. A}\ }\textbf {\bibinfo {volume} {104}},\ \bibinfo
  {pages} {042609} (\bibinfo {year} {2021})}\BibitemShut {NoStop}%
\bibitem [{\citenamefont {Zhang}\ \emph {et~al.}(2022)\citenamefont {Zhang},
  \citenamefont {Chen}, \citenamefont {Bai}, \citenamefont {Wu},\ and\
  \citenamefont {An}}]{PhysRevApplied.17.034073}%
  \BibitemOpen
  \bibfield  {author} {\bibinfo {author} {\bibfnamefont {N.}~\bibnamefont
  {Zhang}}, \bibinfo {author} {\bibfnamefont {C.}~\bibnamefont {Chen}},
  \bibinfo {author} {\bibfnamefont {S.-Y.}\ \bibnamefont {Bai}}, \bibinfo
  {author} {\bibfnamefont {W.}~\bibnamefont {Wu}},\ and\ \bibinfo {author}
  {\bibfnamefont {J.-H.}\ \bibnamefont {An}},\ }\bibfield  {title} {\bibinfo
  {title} {Non-{M}arkovian quantum thermometry},\ }\href
  {https://doi.org/10.1103/PhysRevApplied.17.034073} {\bibfield  {journal}
  {\bibinfo  {journal} {Phys. Rev. Applied}\ }\textbf {\bibinfo {volume}
  {17}},\ \bibinfo {pages} {034073} (\bibinfo {year} {2022})}\BibitemShut
  {NoStop}%
\bibitem [{\citenamefont {Wu}\ and\ \citenamefont
  {An}(2022)}]{PhysRevA.106.062438}%
  \BibitemOpen
  \bibfield  {author} {\bibinfo {author} {\bibfnamefont {W.}~\bibnamefont
  {Wu}}\ and\ \bibinfo {author} {\bibfnamefont {J.-H.}\ \bibnamefont {An}},\
  }\bibfield  {title} {\bibinfo {title} {Quantum speed limit of a noisy
  continuous-variable system},\ }\href
  {https://doi.org/10.1103/PhysRevA.106.062438} {\bibfield  {journal} {\bibinfo
   {journal} {Phys. Rev. A}\ }\textbf {\bibinfo {volume} {106}},\ \bibinfo
  {pages} {062438} (\bibinfo {year} {2022})}\BibitemShut {NoStop}%
\bibitem [{\citenamefont {Ji}\ \emph {et~al.}(2022)\citenamefont {Ji},
  \citenamefont {Bai},\ and\ \citenamefont {An}}]{PhysRevB.106.115427}%
  \BibitemOpen
  \bibfield  {author} {\bibinfo {author} {\bibfnamefont {F.-Z.}\ \bibnamefont
  {Ji}}, \bibinfo {author} {\bibfnamefont {S.-Y.}\ \bibnamefont {Bai}},\ and\
  \bibinfo {author} {\bibfnamefont {J.-H.}\ \bibnamefont {An}},\ }\bibfield
  {title} {\bibinfo {title} {Strong coupling of quantum emitters and the
  exciton polariton in {M}o{S}$_2$ nanodisks},\ }\href
  {https://doi.org/10.1103/PhysRevB.106.115427} {\bibfield  {journal} {\bibinfo
   {journal} {Phys. Rev. B}\ }\textbf {\bibinfo {volume} {106}},\ \bibinfo
  {pages} {115427} (\bibinfo {year} {2022})}\BibitemShut {NoStop}%
\bibitem [{\citenamefont {Wu}\ \emph {et~al.}(2021)\citenamefont {Wu},
  \citenamefont {Bai},\ and\ \citenamefont {An}}]{PhysRevA.103.L010601}%
  \BibitemOpen
  \bibfield  {author} {\bibinfo {author} {\bibfnamefont {W.}~\bibnamefont
  {Wu}}, \bibinfo {author} {\bibfnamefont {S.-Y.}\ \bibnamefont {Bai}},\ and\
  \bibinfo {author} {\bibfnamefont {J.-H.}\ \bibnamefont {An}},\ }\bibfield
  {title} {\bibinfo {title} {Non-{M}arkovian sensing of a quantum reservoir},\
  }\href {https://doi.org/10.1103/PhysRevA.103.L010601} {\bibfield  {journal}
  {\bibinfo  {journal} {Phys. Rev. A}\ }\textbf {\bibinfo {volume} {103}},\
  \bibinfo {pages} {L010601} (\bibinfo {year} {2021})}\BibitemShut {NoStop}%
\bibitem [{\citenamefont {Sheremet}\ \emph {et~al.}(2023)\citenamefont
  {Sheremet}, \citenamefont {Petrov}, \citenamefont {Iorsh}, \citenamefont
  {Poshakinskiy},\ and\ \citenamefont {Poddubny}}]{RevModPhys.95.015002}%
  \BibitemOpen
  \bibfield  {author} {\bibinfo {author} {\bibfnamefont {A.~S.}\ \bibnamefont
  {Sheremet}}, \bibinfo {author} {\bibfnamefont {M.~I.}\ \bibnamefont
  {Petrov}}, \bibinfo {author} {\bibfnamefont {I.~V.}\ \bibnamefont {Iorsh}},
  \bibinfo {author} {\bibfnamefont {A.~V.}\ \bibnamefont {Poshakinskiy}},\ and\
  \bibinfo {author} {\bibfnamefont {A.~N.}\ \bibnamefont {Poddubny}},\
  }\bibfield  {title} {\bibinfo {title} {Waveguide quantum electrodynamics:
  Collective radiance and photon-photon correlations},\ }\href
  {https://doi.org/10.1103/RevModPhys.95.015002} {\bibfield  {journal}
  {\bibinfo  {journal} {Rev. Mod. Phys.}\ }\textbf {\bibinfo {volume} {95}},\
  \bibinfo {pages} {015002} (\bibinfo {year} {2023})}\BibitemShut {NoStop}%
\bibitem [{\citenamefont {Awschalom}\ \emph {et~al.}(2021)\citenamefont
  {Awschalom}, \citenamefont {Berggren}, \citenamefont {Bernien}, \citenamefont
  {Bhave}, \citenamefont {Carr}, \citenamefont {Davids}, \citenamefont
  {Economou}, \citenamefont {Englund}, \citenamefont {Faraon}, \citenamefont
  {Fejer}, \citenamefont {Guha}, \citenamefont {Gustafsson}, \citenamefont
  {Hu}, \citenamefont {Jiang}, \citenamefont {Kim}, \citenamefont {Korzh},
  \citenamefont {Kumar}, \citenamefont {Kwiat}, \citenamefont
  {Lon\ifmmode~\check{c}\else \v{c}\fi{}ar}, \citenamefont {Lukin},
  \citenamefont {Miller}, \citenamefont {Monroe}, \citenamefont {Nam},
  \citenamefont {Narang}, \citenamefont {Orcutt}, \citenamefont {Raymer},
  \citenamefont {Safavi-Naeini}, \citenamefont {Spiropulu}, \citenamefont
  {Srinivasan}, \citenamefont {Sun}, \citenamefont {Vu\ifmmode \check{c}\else
  \v{c}\fi{}kovi\ifmmode~\acute{c}\else \'{c}\fi{}}, \citenamefont {Waks},
  \citenamefont {Walsworth}, \citenamefont {Weiner},\ and\ \citenamefont
  {Zhang}}]{PRXQuantum.2.017002}%
  \BibitemOpen
  \bibfield  {author} {\bibinfo {author} {\bibfnamefont {D.}~\bibnamefont
  {Awschalom}}, \bibinfo {author} {\bibfnamefont {K.~K.}\ \bibnamefont
  {Berggren}}, \bibinfo {author} {\bibfnamefont {H.}~\bibnamefont {Bernien}},
  \bibinfo {author} {\bibfnamefont {S.}~\bibnamefont {Bhave}}, \bibinfo
  {author} {\bibfnamefont {L.~D.}\ \bibnamefont {Carr}}, \bibinfo {author}
  {\bibfnamefont {P.}~\bibnamefont {Davids}}, \bibinfo {author} {\bibfnamefont
  {S.~E.}\ \bibnamefont {Economou}}, \bibinfo {author} {\bibfnamefont
  {D.}~\bibnamefont {Englund}}, \bibinfo {author} {\bibfnamefont
  {A.}~\bibnamefont {Faraon}}, \bibinfo {author} {\bibfnamefont
  {M.}~\bibnamefont {Fejer}}, \bibinfo {author} {\bibfnamefont
  {S.}~\bibnamefont {Guha}}, \bibinfo {author} {\bibfnamefont {M.~V.}\
  \bibnamefont {Gustafsson}}, \bibinfo {author} {\bibfnamefont
  {E.}~\bibnamefont {Hu}}, \bibinfo {author} {\bibfnamefont {L.}~\bibnamefont
  {Jiang}}, \bibinfo {author} {\bibfnamefont {J.}~\bibnamefont {Kim}}, \bibinfo
  {author} {\bibfnamefont {B.}~\bibnamefont {Korzh}}, \bibinfo {author}
  {\bibfnamefont {P.}~\bibnamefont {Kumar}}, \bibinfo {author} {\bibfnamefont
  {P.~G.}\ \bibnamefont {Kwiat}}, \bibinfo {author} {\bibfnamefont
  {M.}~\bibnamefont {Lon\ifmmode~\check{c}\else \v{c}\fi{}ar}}, \bibinfo
  {author} {\bibfnamefont {M.~D.}\ \bibnamefont {Lukin}}, \bibinfo {author}
  {\bibfnamefont {D.~A.}\ \bibnamefont {Miller}}, \bibinfo {author}
  {\bibfnamefont {C.}~\bibnamefont {Monroe}}, \bibinfo {author} {\bibfnamefont
  {S.~W.}\ \bibnamefont {Nam}}, \bibinfo {author} {\bibfnamefont
  {P.}~\bibnamefont {Narang}}, \bibinfo {author} {\bibfnamefont {J.~S.}\
  \bibnamefont {Orcutt}}, \bibinfo {author} {\bibfnamefont {M.~G.}\
  \bibnamefont {Raymer}}, \bibinfo {author} {\bibfnamefont {A.~H.}\
  \bibnamefont {Safavi-Naeini}}, \bibinfo {author} {\bibfnamefont
  {M.}~\bibnamefont {Spiropulu}}, \bibinfo {author} {\bibfnamefont
  {K.}~\bibnamefont {Srinivasan}}, \bibinfo {author} {\bibfnamefont
  {S.}~\bibnamefont {Sun}}, \bibinfo {author} {\bibfnamefont {J.}~\bibnamefont
  {Vu\ifmmode \check{c}\else \v{c}\fi{}kovi\ifmmode~\acute{c}\else
  \'{c}\fi{}}}, \bibinfo {author} {\bibfnamefont {E.}~\bibnamefont {Waks}},
  \bibinfo {author} {\bibfnamefont {R.}~\bibnamefont {Walsworth}}, \bibinfo
  {author} {\bibfnamefont {A.~M.}\ \bibnamefont {Weiner}},\ and\ \bibinfo
  {author} {\bibfnamefont {Z.}~\bibnamefont {Zhang}},\ }\bibfield  {title}
  {\bibinfo {title} {Development of quantum interconnects ({Q}u{IC}s) for
  next-generation information technologies},\ }\href
  {https://doi.org/10.1103/PRXQuantum.2.017002} {\bibfield  {journal} {\bibinfo
   {journal} {PRX Quantum}\ }\textbf {\bibinfo {volume} {2}},\ \bibinfo {pages}
  {017002} (\bibinfo {year} {2021})}\BibitemShut {NoStop}%
\bibitem [{\citenamefont {Lonigro}(2022)}]{Lonigro2022}%
  \BibitemOpen
  \bibfield  {author} {\bibinfo {author} {\bibfnamefont {D.}~\bibnamefont
  {Lonigro}},\ }\bibfield  {title} {\bibinfo {title} {The self-energy of
  friedrichs-lee models and its application to bound states and resonances},\
  }\href {https://doi.org/10.1140/epjp/s13360-022-02690-y} {\bibfield
  {journal} {\bibinfo  {journal} {The European Physical Journal Plus}\ }\textbf
  {\bibinfo {volume} {137}},\ \bibinfo {pages} {492} (\bibinfo {year}
  {2022})}\BibitemShut {NoStop}%
\bibitem [{\citenamefont {Gross}\ and\ \citenamefont
  {Bloch}(2017)}]{doi:10.1126/science.aal3837}%
  \BibitemOpen
  \bibfield  {author} {\bibinfo {author} {\bibfnamefont {C.}~\bibnamefont
  {Gross}}\ and\ \bibinfo {author} {\bibfnamefont {I.}~\bibnamefont {Bloch}},\
  }\bibfield  {title} {\bibinfo {title} {Quantum simulations with ultracold
  atoms in optical lattices},\ }\href {https://doi.org/10.1126/science.aal3837}
  {\bibfield  {journal} {\bibinfo  {journal} {Science}\ }\textbf {\bibinfo
  {volume} {357}},\ \bibinfo {pages} {995} (\bibinfo {year}
  {2017})}\BibitemShut {NoStop}%
\bibitem [{\citenamefont {Yngvesson}(1991)}]{Yngvesson1991}%
  \BibitemOpen
  \bibfield  {author} {\bibinfo {author} {\bibfnamefont {S.}~\bibnamefont
  {Yngvesson}},\ }\bibinfo {title} {Tunneling devices},\ in\ \href
  {https://doi.org/10.1007/978-1-4615-3970-4_4} {\emph {\bibinfo {booktitle}
  {Microwave Semiconductor Devices}}}\ (\bibinfo  {publisher} {Springer US},\
  \bibinfo {address} {Boston, MA},\ \bibinfo {year} {1991})\ pp.\ \bibinfo
  {pages} {103--126}\BibitemShut {NoStop}%
\bibitem [{\citenamefont {Niu}\ \emph {et~al.}(2023)\citenamefont {Niu},
  \citenamefont {Zhang}, \citenamefont {Liu}, \citenamefont {Qiu},
  \citenamefont {Huang}, \citenamefont {Huang}, \citenamefont {Jia},
  \citenamefont {Liu}, \citenamefont {Tao}, \citenamefont {Wei}, \citenamefont
  {Zhou}, \citenamefont {Zou}, \citenamefont {Chen}, \citenamefont {Deng},
  \citenamefont {Deng}, \citenamefont {Hu}, \citenamefont {Hu}, \citenamefont
  {Li}, \citenamefont {Tan}, \citenamefont {Xu}, \citenamefont {Yan},
  \citenamefont {Yan}, \citenamefont {Liu}, \citenamefont {Zhong},
  \citenamefont {Cleland},\ and\ \citenamefont {Yu}}]{Niu2023}%
  \BibitemOpen
  \bibfield  {author} {\bibinfo {author} {\bibfnamefont {J.}~\bibnamefont
  {Niu}}, \bibinfo {author} {\bibfnamefont {L.}~\bibnamefont {Zhang}}, \bibinfo
  {author} {\bibfnamefont {Y.}~\bibnamefont {Liu}}, \bibinfo {author}
  {\bibfnamefont {J.}~\bibnamefont {Qiu}}, \bibinfo {author} {\bibfnamefont
  {W.}~\bibnamefont {Huang}}, \bibinfo {author} {\bibfnamefont
  {J.}~\bibnamefont {Huang}}, \bibinfo {author} {\bibfnamefont
  {H.}~\bibnamefont {Jia}}, \bibinfo {author} {\bibfnamefont {J.}~\bibnamefont
  {Liu}}, \bibinfo {author} {\bibfnamefont {Z.}~\bibnamefont {Tao}}, \bibinfo
  {author} {\bibfnamefont {W.}~\bibnamefont {Wei}}, \bibinfo {author}
  {\bibfnamefont {Y.}~\bibnamefont {Zhou}}, \bibinfo {author} {\bibfnamefont
  {W.}~\bibnamefont {Zou}}, \bibinfo {author} {\bibfnamefont {Y.}~\bibnamefont
  {Chen}}, \bibinfo {author} {\bibfnamefont {X.}~\bibnamefont {Deng}}, \bibinfo
  {author} {\bibfnamefont {X.}~\bibnamefont {Deng}}, \bibinfo {author}
  {\bibfnamefont {C.}~\bibnamefont {Hu}}, \bibinfo {author} {\bibfnamefont
  {L.}~\bibnamefont {Hu}}, \bibinfo {author} {\bibfnamefont {J.}~\bibnamefont
  {Li}}, \bibinfo {author} {\bibfnamefont {D.}~\bibnamefont {Tan}}, \bibinfo
  {author} {\bibfnamefont {Y.}~\bibnamefont {Xu}}, \bibinfo {author}
  {\bibfnamefont {F.}~\bibnamefont {Yan}}, \bibinfo {author} {\bibfnamefont
  {T.}~\bibnamefont {Yan}}, \bibinfo {author} {\bibfnamefont {S.}~\bibnamefont
  {Liu}}, \bibinfo {author} {\bibfnamefont {Y.}~\bibnamefont {Zhong}}, \bibinfo
  {author} {\bibfnamefont {A.~N.}\ \bibnamefont {Cleland}},\ and\ \bibinfo
  {author} {\bibfnamefont {D.}~\bibnamefont {Yu}},\ }\bibfield  {title}
  {\bibinfo {title} {Low-loss interconnects for modular superconducting quantum
  processors},\ }\href {https://doi.org/10.1038/s41928-023-00925-z} {\bibfield
  {journal} {\bibinfo  {journal} {Nature Electronics}\ }\textbf {\bibinfo
  {volume} {6}},\ \bibinfo {pages} {235} (\bibinfo {year} {2023})}\BibitemShut
  {NoStop}%
\bibitem [{\citenamefont {Lei}\ \emph {et~al.}(2023)\citenamefont {Lei},
  \citenamefont {Fukumori}, \citenamefont {Rochman}, \citenamefont {Zhu},
  \citenamefont {Endres}, \citenamefont {Choi},\ and\ \citenamefont
  {Faraon}}]{Lei2023}%
  \BibitemOpen
  \bibfield  {author} {\bibinfo {author} {\bibfnamefont {M.}~\bibnamefont
  {Lei}}, \bibinfo {author} {\bibfnamefont {R.}~\bibnamefont {Fukumori}},
  \bibinfo {author} {\bibfnamefont {J.}~\bibnamefont {Rochman}}, \bibinfo
  {author} {\bibfnamefont {B.}~\bibnamefont {Zhu}}, \bibinfo {author}
  {\bibfnamefont {M.}~\bibnamefont {Endres}}, \bibinfo {author} {\bibfnamefont
  {J.}~\bibnamefont {Choi}},\ and\ \bibinfo {author} {\bibfnamefont
  {A.}~\bibnamefont {Faraon}},\ }\bibfield  {title} {\bibinfo {title}
  {Many-body cavity quantum electrodynamics with driven inhomogeneous
  emitters},\ }\href {https://doi.org/10.1038/s41586-023-05884-1} {\bibfield
  {journal} {\bibinfo  {journal} {Nature}\ }\textbf {\bibinfo {volume} {617}},\
  \bibinfo {pages} {271} (\bibinfo {year} {2023})}\BibitemShut {NoStop}%
\bibitem [{\citenamefont {Mirhosseini}\ \emph {et~al.}(2019)\citenamefont
  {Mirhosseini}, \citenamefont {Kim}, \citenamefont {Zhang}, \citenamefont
  {Sipahigil}, \citenamefont {Dieterle}, \citenamefont {Keller}, \citenamefont
  {Asenjo-Garcia}, \citenamefont {Chang},\ and\ \citenamefont
  {Painter}}]{Mirhosseini2019}%
  \BibitemOpen
  \bibfield  {author} {\bibinfo {author} {\bibfnamefont {M.}~\bibnamefont
  {Mirhosseini}}, \bibinfo {author} {\bibfnamefont {E.}~\bibnamefont {Kim}},
  \bibinfo {author} {\bibfnamefont {X.}~\bibnamefont {Zhang}}, \bibinfo
  {author} {\bibfnamefont {A.}~\bibnamefont {Sipahigil}}, \bibinfo {author}
  {\bibfnamefont {P.~B.}\ \bibnamefont {Dieterle}}, \bibinfo {author}
  {\bibfnamefont {A.~J.}\ \bibnamefont {Keller}}, \bibinfo {author}
  {\bibfnamefont {A.}~\bibnamefont {Asenjo-Garcia}}, \bibinfo {author}
  {\bibfnamefont {D.~E.}\ \bibnamefont {Chang}},\ and\ \bibinfo {author}
  {\bibfnamefont {O.}~\bibnamefont {Painter}},\ }\bibfield  {title} {\bibinfo
  {title} {Cavity quantum electrodynamics with atom-like mirrors},\ }\href
  {https://doi.org/10.1038/s41586-019-1196-1} {\bibfield  {journal} {\bibinfo
  {journal} {Nature}\ }\textbf {\bibinfo {volume} {569}},\ \bibinfo {pages}
  {692} (\bibinfo {year} {2019})}\BibitemShut {NoStop}%
\bibitem [{\citenamefont {Blais}\ \emph {et~al.}(2020)\citenamefont {Blais},
  \citenamefont {Girvin},\ and\ \citenamefont {Oliver}}]{Blais2020}%
  \BibitemOpen
  \bibfield  {author} {\bibinfo {author} {\bibfnamefont {A.}~\bibnamefont
  {Blais}}, \bibinfo {author} {\bibfnamefont {S.~M.}\ \bibnamefont {Girvin}},\
  and\ \bibinfo {author} {\bibfnamefont {W.~D.}\ \bibnamefont {Oliver}},\
  }\bibfield  {title} {\bibinfo {title} {Quantum information processing and
  quantum optics with circuit quantum electrodynamics},\ }\href
  {https://doi.org/10.1038/s41567-020-0806-z} {\bibfield  {journal} {\bibinfo
  {journal} {Nature Physics}\ }\textbf {\bibinfo {volume} {16}},\ \bibinfo
  {pages} {247} (\bibinfo {year} {2020})}\BibitemShut {NoStop}%
\bibitem [{\citenamefont {Blais}\ \emph {et~al.}(2021)\citenamefont {Blais},
  \citenamefont {Grimsmo}, \citenamefont {Girvin},\ and\ \citenamefont
  {Wallraff}}]{RevModPhys.93.025005}%
  \BibitemOpen
  \bibfield  {author} {\bibinfo {author} {\bibfnamefont {A.}~\bibnamefont
  {Blais}}, \bibinfo {author} {\bibfnamefont {A.~L.}\ \bibnamefont {Grimsmo}},
  \bibinfo {author} {\bibfnamefont {S.~M.}\ \bibnamefont {Girvin}},\ and\
  \bibinfo {author} {\bibfnamefont {A.}~\bibnamefont {Wallraff}},\ }\bibfield
  {title} {\bibinfo {title} {Circuit quantum electrodynamics},\ }\href
  {https://doi.org/10.1103/RevModPhys.93.025005} {\bibfield  {journal}
  {\bibinfo  {journal} {Rev. Mod. Phys.}\ }\textbf {\bibinfo {volume} {93}},\
  \bibinfo {pages} {025005} (\bibinfo {year} {2021})}\BibitemShut {NoStop}%
\bibitem [{\citenamefont {Song}\ \emph {et~al.}(2024)\citenamefont {Song},
  \citenamefont {Liu}, \citenamefont {Zhou}, \citenamefont {Yang},\ and\
  \citenamefont {An}}]{PhysRevLett.132.090401}%
  \BibitemOpen
  \bibfield  {author} {\bibinfo {author} {\bibfnamefont {W.-L.}\ \bibnamefont
  {Song}}, \bibinfo {author} {\bibfnamefont {H.-B.}\ \bibnamefont {Liu}},
  \bibinfo {author} {\bibfnamefont {B.}~\bibnamefont {Zhou}}, \bibinfo {author}
  {\bibfnamefont {W.-L.}\ \bibnamefont {Yang}},\ and\ \bibinfo {author}
  {\bibfnamefont {J.-H.}\ \bibnamefont {An}},\ }\bibfield  {title} {\bibinfo
  {title} {Remote charging and degradation suppression for the quantum
  battery},\ }\href {https://doi.org/10.1103/PhysRevLett.132.090401} {\bibfield
   {journal} {\bibinfo  {journal} {Phys. Rev. Lett.}\ }\textbf {\bibinfo
  {volume} {132}},\ \bibinfo {pages} {090401} (\bibinfo {year}
  {2024})}\BibitemShut {NoStop}%
\bibitem [{\citenamefont {Bai}\ and\ \citenamefont
  {An}(2021)}]{PhysRevLett.127.083602}%
  \BibitemOpen
  \bibfield  {author} {\bibinfo {author} {\bibfnamefont {S.-Y.}\ \bibnamefont
  {Bai}}\ and\ \bibinfo {author} {\bibfnamefont {J.-H.}\ \bibnamefont {An}},\
  }\bibfield  {title} {\bibinfo {title} {Generating stable spin squeezing by
  squeezed-reservoir engineering},\ }\href
  {https://doi.org/10.1103/PhysRevLett.127.083602} {\bibfield  {journal}
  {\bibinfo  {journal} {Phys. Rev. Lett.}\ }\textbf {\bibinfo {volume} {127}},\
  \bibinfo {pages} {083602} (\bibinfo {year} {2021})}\BibitemShut {NoStop}%
\end{thebibliography}
\end{document}